\documentclass[amsmath,amssymb,aps,prb,twocolumn,nolongbibliography,floatfix]{revtex4-2}
\usepackage{graphicx}
\usepackage{bm}
\usepackage{physics}
\usepackage{booktabs}
\usepackage{xcolor}
\usepackage{hyperref}
\usepackage{orcidlink}

\begin{document}

\title{Orbital Angular Momentum Currents and Textures in a Multiorbital Double Helix}

\author{Danny Cordova\orcidlink{0009-0005-5675-9395}}
\affiliation{Departamento de F\'isica, Colegio de Ciencias e Ingenier\'ia, Universidad San Francisco de Quito, Diego de Robles y Via Interoce\'anica, Quito 17901, Ecuador}
\author{Edwin V\'asconez\orcidlink{0009-0005-8575-0226}}
\affiliation{Departamento de F\'isica, Colegio de Ciencias e Ingenier\'ia, Universidad San Francisco de Quito, Diego de Robles y Via Interoce\'anica, Quito 17901, Ecuador}
\author{Ernesto Medina\orcidlink{0000-0002-1566-0170}}
\email[Corresponding author: ]{emedina@usfq.edu.ec}
\affiliation{Departamento de F\'isica, Colegio de Ciencias e Ingenier\'ia, Universidad San Francisco de Quito, Diego de Robles y Via Interoce\'anica, Quito 17901, Ecuador}

\begin{abstract}

The microscopic origin of chirality-induced spin selectivity (CISS) in
biomolecular structures has traditionally been discussed in terms of
spin--orbit coupling and spin angular momentum.
However, orbital angular momentum (OAM) is increasingly recognized as a more direct electronic response to molecular chirality.
A helical
hopping network built only from $s$ orbitals can support an itinerant
orbital moment and an orbital Edelstein response, but such a basis has
no local atomic OAM and therefore omits the complementary contribution
arising from coherence among active $p$ orbitals. Here we construct a
six-band multiorbital double-helix model parameterized by characteristic
B-form DNA geometry and hopping scales. The model incorporates
major--minor groove asymmetry, an axial base-pair offset representing
stagger, and compositional inequivalence between the two strands,
providing structural control parameters with which the orbital response
can be tuned. Inter-strand hybridization generates a full
three-component atomic OAM texture in the co-rotating local helical
frame, including a radial component absent in the corresponding
single-helix model. We separately evaluate the itinerant OAM using the
Bloch-wave-packet formulation of orbital magnetization and identify an
additional inter-strand circulation channel unique to the double-helix
connectivity. The resulting atomic and itinerant responses have
distinct microscopic origins, magnitudes, and temperature dependences,
establishing a multiorbital framework for geometry-controlled angular-momentum generation and transport in chiral molecular conductors.

\end{abstract}

\maketitle

\section{Introduction}

Chirality-induced spin selectivity (CISS) has revealed sizeable spin-dependent responses in DNA, helical proteins, molecular layers, and other chiral structures built largely from light, nonmagnetic elements \cite{reviewPaltiel,bloom2024chiral,aiello2022chirality,eckvahl2024detecting}. Microscopic descriptions commonly combine structural chirality with spin--orbit coupling (SOC) \cite{varela2016effective,gutierrez2015spin,guo2014spin}, together with contacts, dephasing, or other nonequilibrium ingredients needed to evade reciprocity constraints in ideal coherent two-terminal transport \cite{yang2020detecting,yang2020connecting}. The comparatively weak atomic SOC of organic molecules and the difficulty of accounting for the magnitude and diversity of CISS signals \cite{varela2024optical} motivate a complementary question: how strongly can chirality act first on the orbital degree of freedom?

Orbital angular momentum (OAM) can respond directly to structural chirality without requiring atomic SOC. Yoda, Yokoyama, and Murakami showed that chiral electronic structures can develop current-induced orbital magnetization, or an orbital Edelstein response \cite{Yoda2015HelicalMagnetization,Yoda2018OrbitalEdelstein}. 
More generally, orbital textures associated with hybridization,
crystal symmetry, and the geometrical structure of Bloch states have
emerged as central ingredients of orbitronics, with both intra-atomic
and inter-site contributions playing distinct roles
\cite{Go2018OrbitalTexture,Cysne2025Orbitronics}.
Such orbital responses may subsequently be converted into spin
polarization through SOC or through interfacial orbital-to-spin
conversion, providing a possible intermediate stage between structural
chirality and the experimentally detected spin response \cite{varela2023spin,ozeri2023modification,mishra2020spin}.

Two microscopic OAM mechanisms should be distinguished in a helical conductor. The \emph{itinerant} contribution is the self-rotation of a propagating Bloch wave packet through the real-space hopping network. G\"obel, Schimpf, and Mertig showed that this mechanism alone produces an orbital moment and orbital Edelstein response in a helical model containing only $s$ orbitals \cite{GobelSchimpfMertig2025}; related first-principles work finds substantial orbital textures in DNA-like chiral structures \cite{Liu2021ChiralityTopological}. A complementary \emph{atomic} contribution appears when the local basis itself carries OAM. In a $p$-orbital helix, Slater--Koster hybridization among differently oriented local orbitals generates momentum-dependent inter-orbital coherences and hence an on-site OAM texture, as also emphasized in multiorbital helical-chain transport \cite{YatabeAkera2023} and in recent analyses of chirality-encoded molecular wave functions \cite{Georgiou2025Chirality}.

In our previous single-helix model \cite{CordovaBercheMedina2026}, a minimal local basis $(p_r,p_\phi,p_z)$ generated atomic OAM without atomic SOC. The azimuthal and longitudinal textures were finite, whereas the radial component vanished because of the phase structure of the single-strand Bloch Hamiltonian. Time-reversal symmetry forced the equilibrium OAM accumulation to vanish but allowed a conventional equilibrium OAM current; under a longitudinal field, the same odd-in-$k$ texture produced a linear orbital Edelstein polarization while the first-order field-induced OAM current vanished by parity.

Here we extend that framework to an asymmetric B-DNA-inspired double helix. The second strand introduces inter-strand hybridization, strand inequivalence, major--minor groove asymmetry, axial stagger, and new closed inter-site circulation paths. We construct a six-band $p$-orbital tight-binding Hamiltonian and show that inter-strand hybridization releases the single-helix phase constraint, producing a full three-component atomic OAM texture with a finite radial component. We then determine the equilibrium conventional OAM current and the driven response through second order in a longitudinal electric field: a linear atomic orbital Edelstein polarization is accompanied by a quadratic field-induced OAM current. Separately, we evaluate the itinerant OAM with the Bloch-wave-packet formulation of orbital magnetization, calculate its longitudinal texture and linear and quadratic responses, and identify a double-helix plaquette with a purely radial vector area as an additional transverse itinerant channel. The model therefore places atomic and itinerant OAM on the same microscopic footing while keeping their distinct physical origins explicit.

\section{Tight-Binding Hamiltonian Framework}
\label{section_tight-binding}

\subsection{Slater--Koster $p$-Orbital Framework}

We represent the electronic structure of the double helix by a minimal orthogonal basis of three local $p$-like orbitals ($|p_r\rangle, |p_\phi\rangle, |p_z\rangle$) associated with each coarse-grained nucleotide site. The kinetic charge transfer between any two sites in the lattice is governed by the generalized Slater--Koster framework. For an arbitrary bond vector $\mathbf{d}$ connecting two sites, possessing a unit direction $\hat{\mathbf{d}} = \mathbf{d}/|\mathbf{d}|$, the transfer integral $t_{ij}$ between a local $p$-orbital oriented along $\hat{\mathbf{e}}_i$ and another along $\hat{\mathbf{e}}_j$ is dictated by the fundamental overlap integrals $V_{pp\sigma}$ and $V_{pp\pi}$:
\begin{equation}
t_{ij} = V_{pp\pi} (\hat{\mathbf{e}}_i \cdot \hat{\mathbf{e}}_j) + (V_{pp\sigma} - V_{pp\pi}) (\hat{\mathbf{e}}_i \cdot \hat{\mathbf{d}})(\hat{\mathbf{e}}_j \cdot \hat{\mathbf{d}}).
\end{equation}
This formulation yields the effective transfer matrices used below. The geometrical parameters are chosen from characteristic B-form DNA structural scales, while the on-site energies and hopping amplitudes constitute a representative tight-binding parameterization consistent with the energy scales used in model descriptions of DNA charge transport \cite{hawke2010electronic, varela2016effective}. Table~\ref{tab:bdna_parameters} in the appendix separates the structural input from the model's effective electronic parameters. We derive their geometrical dependence in the subsequent sections.

\subsection{Double-Helix Geometry and Vertical Shear}

The structural scaffolding of B-form DNA is parameterized as two discrete, intertwined helices identified by the strand index $\mu \in \{1,2\}$ and the sequential nucleotide index $n$. In a global Cartesian frame $(X, Y, Z)$, the spatial coordinates of the lattice sites are defined as:
\begin{align}
\mathbf{R}_{1,n} &= a \cos(n\Delta\varphi)\mathbf{\hat{X}} + a \sin(n\Delta\varphi)\mathbf{\hat{Y}} + n h_z \mathbf{\hat{Z}}, \\
\mathbf{R}_{2,n} &= a \cos(n\Delta\varphi + \varphi_0)\mathbf{\hat{X}} + a \sin(n\Delta\varphi + \varphi_0)\mathbf{\hat{Y}} \nonumber \\
&\quad+ (n h_z + z_0) \mathbf{\hat{Z}},
\end{align}
where $a$ defines the helix radius, $h_z$ denotes the axial rise per nucleotide, and $\Delta\varphi$ represents the intra-strand azimuthal twist angle. The phase angle $\varphi_0 \approx 150^\circ$ governs the relative azimuthal separation of the second strand, inducing the asymmetric formation of the major and minor grooves characteristic of B-DNA. The parameter $z_0$ strictly defines the longitudinal displacement between the two complementary bases at a single rung. While simplified models often assume $ z_0=0$, realistic DNA geometries in solution exhibit a finite vertical shear, crystallographically termed base-pair stagger. This vertical shear deviates from strict planarity, possessing an average characteristic value of $z_0 \approx 0.15$ {\AA} \cite{shui1998}.

\begin{figure}[t]
    \centering
    \includegraphics[width=0.8 \linewidth]{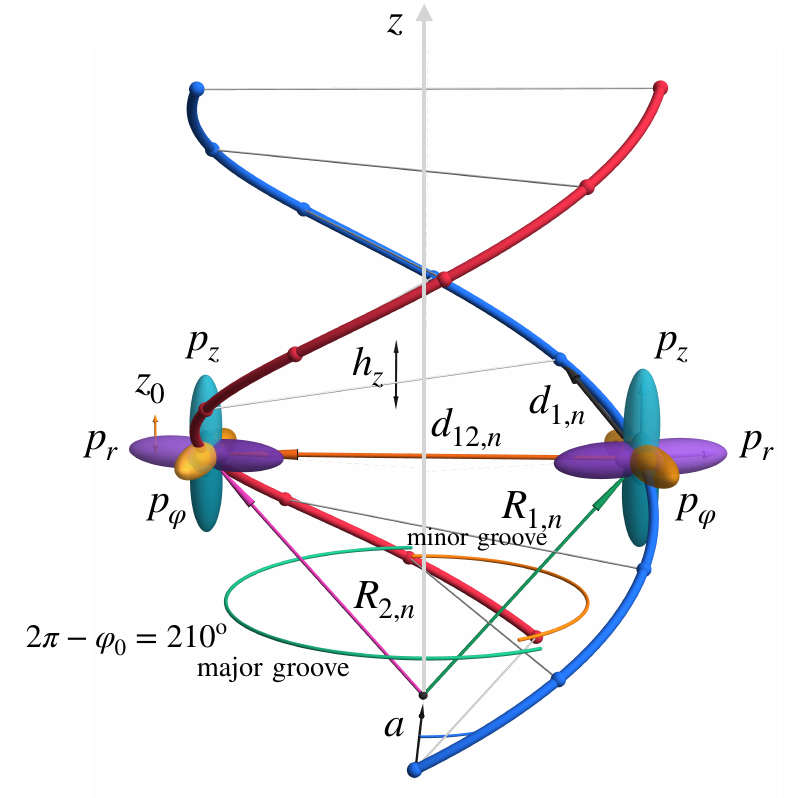}
    \caption{Schematic representation of the double helix geometry and local orbital basis. The red and blue backbones represent the two helical strands wrapping around the central $z$-axis, with radius $a$ and axial rise $h_z$. The structural asymmetry is defined by the minor groove angle $\varphi_0$ and the major groove angle $2\pi - \varphi_0 = 210^\circ$, as well as an inter-strand vertical shear $z_0$. Position vectors $\mathbf R_{1,n}$ and $\mathbf R_{2,n}$ locate site $n$ on each strand, with $\mathbf d_{1,n}$ and $\mathbf d_{12,n}$ representing the intra-strand and inter-strand bond vectors, respectively. At each lattice site, the local electronic states are modeled using $p_r$, $p_\phi$, and $p_z$ orbitals.}
    \label{fig:model}
\end{figure}

To accurately evaluate the transfer integrals, a localized, co-rotating cylindrical coordinate frame must be established at every discrete site $(n, \mu)$ to orient the minimal orthogonal orbital basis $\{|p_r\rangle, |p_\phi\rangle, |p_z\rangle\}$:
\begin{align}
\mathbf{\hat{r}}_{\mu,n} &= \cos(n\Delta\varphi + \delta_\mu)\mathbf{\hat{X}} + \sin(n\Delta\varphi + \delta_\mu)\mathbf{\hat{Y}}, \\
\mathbf{\hat{\phi}}_{\mu,n} &= -\sin(n\Delta\varphi + \delta_\mu)\mathbf{\hat{X}} + \cos(n\Delta\varphi + \delta_\mu)\mathbf{\hat{Y}}, \\
\mathbf{\hat{z}}_{\mu,n} &= \mathbf{\hat{Z}},
\end{align}
where $\delta_1 = 0$ and $\delta_2 = \varphi_0$. This local rotational mapping ensures that the resulting Slater--Koster hybridizations encode the molecular geometry. We denote the intra-strand bond by $\mathbf d_{\mu,n}=\mathbf R_{\mu,n+1}-\mathbf R_{\mu,n}$ and the inter-strand rung by $\mathbf d_{12,n}=\mathbf R_{2,n}-\mathbf R_{1,n}$. The model is represented in Fig.~\ref{fig:model}.
Figure~\ref{fig:model} summarizes the geometrical parameters and the co-rotating local $p$-orbital basis used throughout the tight-binding construction.
The intra-strand nearest-neighbor distance is $d = \sqrt{4a^2 \sin^2(\Delta\varphi/2) + h_z^2}$. The unit bond vector connecting adjacent intra-strand sites $\hat{\mathbf{d}}_n = (\mathbf{R}_{\mu,n+1} - \mathbf{R}_{\mu,n})/d$ yields specific projections onto the local basis, generating the direction cosines $\alpha = -\hat{\mathbf{r}}_{\mu, n+1} \cdot \hat{\mathbf{d}}_{n}$, $\beta = \hat{\bm{\phi}}_{\mu, n+1} \cdot \hat{\mathbf{d}}_{n}$, and $\gamma = \mathbf{\hat{z}} \cdot \hat{\mathbf{d}}_{n}$.

\subsection{Inter-Strand Coupling and Structural Asymmetry}

The two complementary strands are bridged by the inter-strand coupling matrix $T^{\text{inter}}$. The rung vector $\mathbf d_{12,n}$ has a transverse magnitude $d_\perp = 2a\sin(\varphi_0/2)$ and, due to the vertical shear, a longitudinal component $z_0$. Its length is $d_{\rm int}=|\mathbf d_{12,n}|=\sqrt{d_\perp^2+z_0^2}$.

The direction cosine characterizing the out-of-plane tilt is $\gamma_{\text{int}}=z_0/d_{\rm int}$. We determine the projections of the normalized vector $\hat{\mathbf d}_{12,n}$ onto the local bases of strand 1 and strand 2. Substituting these geometric overlaps into the inter-strand Slater--Koster parameters yields the fully coupled $3 \times 3$ coupling block:
\begin{equation}
T^{\text{inter}} = \begin{pmatrix}
t_{\perp, rr} & t_{\perp, r\phi} & t_{\perp, rz} \\
t_{\perp, \phi r} & t_{\perp, \phi\phi} & t_{\perp, \phi z} \\
t_{\perp, zr} & t_{\perp, z\phi} & t_{\perp, zz}
\end{pmatrix}.
\end{equation}
The purely diagonal elements are analytically formulated as:
\begin{align}
t_{\perp, rr} &= V_{pp\pi}^\perp \cos\varphi_0 - \Delta^\perp \sin^2(\varphi_0/2)(1-\gamma_{\text{int}}^2), \\
t_{\perp, \phi\phi} &= V_{pp\pi}^\perp \cos\varphi_0 + \Delta^\perp \cos^2(\varphi_0/2)(1-\gamma_{\text{int}}^2), \\
t_{\perp, zz} &= V_{pp\pi}^\perp + \Delta^\perp \gamma_{\text{int}}^2,
\end{align}
where $\Delta^{\perp}=V_{pp\sigma}^{\perp}-V_{pp\pi}^{\perp}$, while $V^\perp_{pp\sigma}$ and $V^\perp_{pp\pi}$ are the inter-strand $\sigma$- and $\pi$-type Slater--Koster integrals.
The in-plane structural asymmetry dictated by the major and minor groove offset ($\varphi_0 \neq \pi$) defines the transverse cross-couplings:
\begin{align}
t_{\perp, r\phi} &= -\sin\varphi_0 \left[V_{pp\pi}^{\perp}+\frac{\Delta^{\perp}}{2}(1-\gamma_{\rm{int}}^2)\right], \\
t_{\perp, \phi r} &= -t_{\perp, r\phi}.
\end{align}
The presence of the vertical shear ($z_0 \neq 0$) activates out-of-plane cross-couplings between the $p_z$ orbitals and the transverse $p_r$ and $p_\phi$ orbitals across the gap:
\begin{align}
t_{\perp, rz} &= -\Delta^\perp \sin(\varphi_0/2) \sqrt{1-\gamma_{\text{int}}^2} \gamma_{\text{int}}, \\
t_{\perp, \phi z} &= \Delta^\perp \cos(\varphi_0/2) \sqrt{1-\gamma_{\text{int}}^2} \gamma_{\text{int}},\\
t_{\perp, zr} &= -t_{\perp, rz},\\
t_{\perp, z\phi} &= -t_{\perp, \phi z}.
\end{align}
Consequently, the matrix representing hopping from strand 1 to strand 2 is not symmetric in its local-orbital indices, $(T^{\text{inter}})^T \neq T^{\text{inter}}$. Hermiticity of the full Hamiltonian is nevertheless preserved because the reverse hopping block is $(T^{\text{inter}})^\dagger$.

\section{Energy Spectrum and Analytical Eigenfunctions} \label{sec:energies}

\subsection{Intra-Strand Kinetics and Compositional Asymmetry}

We construct the intra-strand Hamiltonian governing electron propagation along a single helical backbone by considering a tight-binding framework based on local $p$-orbitals ($\hat{\mathbf{r}}, \hat{\bm{\phi}}, \hat{\mathbf{z}}$). Following the Slater--Koster parameterization developed for helical architectures in Ref.~\cite{CordovaBercheMedina2026}, the intra-strand hopping parameters account for spatial chirality through a coordinate transformation that links neighboring sites across a step height $h_z$ and turning angle $\Delta\varphi$. Because B-DNA is composed of two distinct nucleotide sequences forming an asymmetric double helix, the on-site atomic energies of the two strands are inherently different. We define the strand-dependent on-site energies as $\epsilon_{\mu, \alpha}^0$ for orbital $\alpha \in \{r, \phi, z\}$ on strand $\mu \in \{1, 2\}$. Applying the Bloch ansatz $c_{k,\mu} = \frac{1}{\sqrt{N}} \sum_n e^{i k n h_z} c_{n,\mu}$,
where \(N\) is the number of unit cells and \(c_{n,\mu}=(c_{n,\mu,r},c_{n,\mu,\phi},c_{n,\mu,z})^T\) denotes the local orbital spinor, we obtain two distinct $3 \times 3$ Bloch matrices, $H_1(k)$ and $H_2(k)$, for the isolated strands:
\begin{equation}
H_\mu(k) = \begin{pmatrix}
\epsilon_{\mu,r}(k) & i \gamma_{r\phi}(k) & i \gamma_{rz}(k) \\
-i \gamma_{r\phi}(k) & \epsilon_{\mu,\phi}(k) & \gamma_{\phi z}(k) \\
-i \gamma_{rz}(k) & \gamma_{\phi z}(k) & \epsilon_{\mu,z}(k)
\end{pmatrix}.
\label{eq:H_k}
\end{equation}
Spatial chirality manifests through the purely imaginary, odd-parity momentum hybridization channels $\gamma_{r\phi}(k)$ and $\gamma_{rz}(k)$, which are identical for both strands due to their matching local rotational frames, while the real coupling $\gamma_{\phi z}(k)$ maintains even parity under momentum inversion. The complete analytical definitions of these matrix elements and the underlying Slater--Koster parameters are detailed in Appendix~\ref{app:slater_koster}.

\subsection{Bonding--Antibonding Representation of the 6-Band Hamiltonian}

The total system requires the full $6 \times 6$ Bloch matrix formed by the two structurally and compositionally asymmetric strands:
\begin{equation}
\mathcal{H}(k) = \begin{pmatrix} H_1(k) & T^{\text{inter}} \\ (T^{\text{inter}})^\dagger & H_2(k) \end{pmatrix}, \label{eq:H6x6}
\end{equation}
where $T^{\text{inter}}$ is the inter-strand hopping matrix. To expose the bonding--antibonding structure, we apply the unitary transformation $V = \frac{1}{\sqrt{2}} \begin{pmatrix} \mathbb{I}_3 & \mathbb{I}_3 \\ \mathbb{I}_3 & -\mathbb{I}_3 \end{pmatrix}$, where $\mathbb{I}_3$ is the $3\times3$ identity matrix. This gives
\begin{equation}
\begin{aligned}
V^\dagger \mathcal{H} V &= \frac{1}{2} \begin{pmatrix} \mathbb{I}_3 & \mathbb{I}_3 \\ \mathbb{I}_3 & -\mathbb{I}_3 \end{pmatrix} \begin{pmatrix} H_1 + T^{\text{inter}} & H_1 - T^{\text{inter}} \\ (T^{\text{inter}})^\dagger + H_2 & (T^{\text{inter}})^\dagger - H_2 \end{pmatrix} \\
&= \begin{pmatrix} \bar{H} + T_S & \frac{1}{2}\Delta H - T_A \\ \frac{1}{2}\Delta H + T_A & \bar{H} - T_S \end{pmatrix},
\label{eq:V_transformed}
\end{aligned}
\end{equation}
where $\bar H=(H_1+H_2)/2$, $\Delta H=H_1-H_2$, $T_S=[T^{\rm inter}+(T^{\rm inter})^\dagger]/2$, and $T_A=[T^{\rm inter}-(T^{\rm inter})^\dagger]/2$.
Equation~\eqref{eq:V_transformed} expresses the Hamiltonian in the bonding--antibonding basis, but it does not, in general, block-diagonalize the system. The diagonal blocks, $\bar{H}+T_S$ and $\bar{H}-T_S$, would define independent bonding ($+$) and antibonding ($-$) sectors only if the two off-diagonal blocks vanished.
Exact decoupling, therefore, requires simultaneously
$\Delta H = 0, T_A = 0$. The first condition corresponds to two equivalent strands, $H_1=H_2$, whereas the second requires a Hermitian inter-strand coupling matrix,
$T^{\mathrm{inter}}=\left(T^{\mathrm{inter}}\right)^\dagger$.
In the present double-helix model, strand inequivalence gives $\Delta H\neq 0$, while the asymmetric cross-orbital inter-strand hoppings generated by the double-helix geometry produce $T_A\neq 0$. Consequently, the bonding and antibonding sectors remain coupled, and the labels $+$ and $-$ are not exact quantum numbers of the full Hamiltonian. They may still characterize the dominant bonding or antibonding content of individual eigenstates when the corresponding mixing is weak.

The strength of the bonding--antibonding mixing cannot be inferred solely from the fact that $\Delta H$ and $T_A$ are nonzero. The magnitude of the off-diagonal matrix elements relative to the energy separation between the corresponding states controls it. The orbital and bonding character of the exact six-band eigenstates must therefore be determined directly from their eigenvector weights.

\subsection{Characteristic Equation and Adjugate Spinors}

\begin{figure}[t]
    \centering
    \includegraphics[width=1\linewidth]{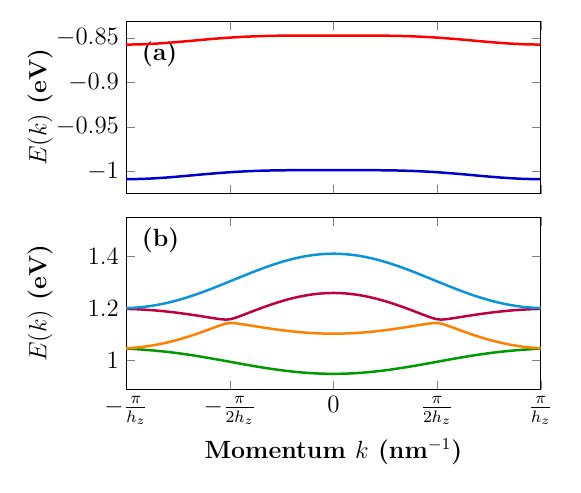}
    \caption{Electronic band structure of the fully coupled six-band B-DNA double-helix model. Panel (a) isolates the lowest energy states: $\nu=1$ (blue), $\nu=2$ (red), corresponding primarily to the longitudinal $p_z$ orbitals. Panel (b) details the remaining higher-energy manifolds: $\nu=3$ (green), $\nu=4$ (orange), $\nu=5$ (purple), $\nu=6$ (cyan), dominated by hybridized transverse modes. The structural and compositional asymmetry lifts the degeneracies of the corresponding symmetric model.}
    \label{fig:band_structure}
\end{figure}

Because the system cannot be decoupled, we must evaluate the full sixth-order characteristic secular equation:
\begin{equation}
\det(\mathcal{H}(k) - E\mathbb{I}_6) = 0.
\end{equation}
Expanding this determinant yields a sixth-degree polynomial in energy $P_6(E, k) = 0$. The exact analytical energy bands $E_\nu(k)$ correspond to the six real roots of this polynomial, which are numerically plotted in Fig.~\ref{fig:band_structure}.

The derived energy spectrum, illustrated in Fig.~\ref{fig:band_structure}, demonstrates the impact of the coupled structural and compositional asymmetries. Because $\Delta H \neq 0$ and $T_A \neq 0$, bonding and antibonding character is mixed rather than defining two exact, independent sectors. The resulting bands show avoided crossings and lifted orbital degeneracies. The lower-energy manifold in panel (a) is energetically separated due to the deep on-site energy of the longitudinal mode, while the transverse modes in panel (b) hybridize heavily due to the chiral inter-strand overlap. For each exact energy eigenvalue $E_\nu(k)$, the corresponding six-component analytical Bloch spinor $|\Psi_\nu(k)\rangle$ is formally defined using the algebraic adjugate matrix formulation. The unnormalized spinor components $\Psi_{j,\nu}(k)$ correspond to any nonzero column of the classical adjoint matrix $\text{Adj}(\mathcal{H}(k) - E_\nu(k)\mathbb{I}_6)$. Denoting $M_{i,j}$ as the $(i,j)$ minor of the characteristic matrix, the eigenvector components are analytically expressed as:
\begin{equation}
\Psi_{j,\nu}(k) \propto (-1)^{i+j} \det \left[ M_{i,j}(\mathcal{H}(k) - E_\nu(k)\mathbb{I}_6) \right],
\end{equation}
where $i$ is a fixed row index chosen such that the resulting vector is nontrivial. The normalized eigenfunctions are subsequently obtained by dividing by the vector norm, yielding the local orbital amplitudes $u_{1,r}, u_{1,\phi}, u_{1,z}$ for the first strand and $u_{2,r}, u_{2,\phi}, u_{2,z}$ for the second strand.

\section{Atomic Orbital Angular Momentum Textures}
\label{sec:atomic_oam}

The first contribution considered is the local, or atomic, orbital angular momentum associated with the on-site $p$-orbital manifold. This quantity is defined by the usual angular-momentum operators acting within the local basis $(p_r,p_\phi,p_z)$ on each strand and is distinct from the itinerant contribution associated with inter-site wave-packet circulation, which will be treated separately in Sec.~\ref{sec:itinerant_oam}.

\subsection{Local Atomic OAM Operators and Textures}

\begin{figure}[t]
    \centering
    \includegraphics[width=0.9\linewidth]{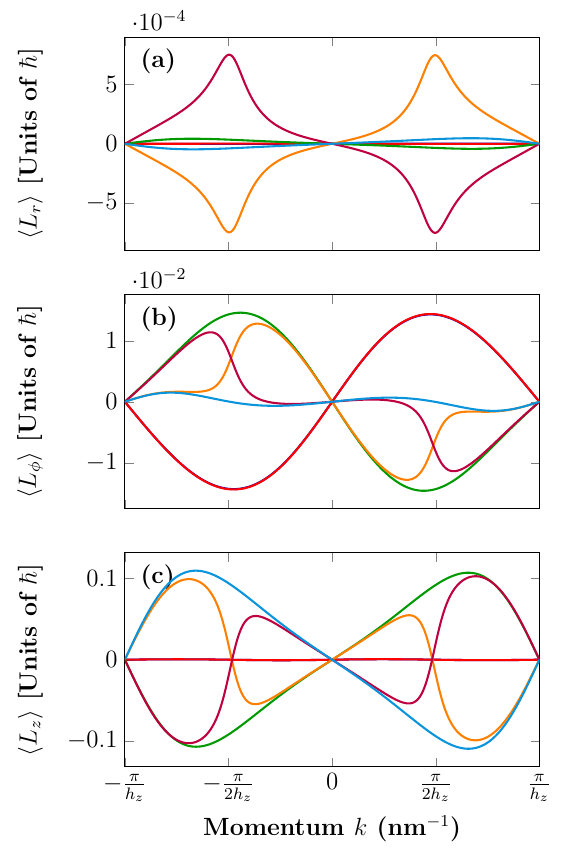}
    \caption{ Momentum-dependent orbital angular momentum textures for the six-band double-helix model. The color coding follows the band indexing ($\nu=1, \dots, 6$) established in Fig.~\ref{fig:band_structure}. Panels (a), (b), and (c) display the radial, azimuthal, and longitudinal components, respectively. Time-reversal symmetry imposes the antisymmetric orbital texture $\langle L_\alpha\rangle_\nu(k)=-\langle L_\alpha\rangle_\nu(-k)$. Crucially, panel (a) demonstrates the emergence of a finite radial orbital texture $\langle L_r\rangle_\nu(k)$, which is absent in the corresponding isolated single-helix model.
    }
    \label{fig:textures}
\end{figure}

We compute the exact momentum-dependent local atomic OAM textures for band $\nu$, defined generally as $\langle L_\alpha \rangle_{\nu}(k) = \langle \Psi_\nu(k) | \hat{L}_\alpha | \Psi_\nu(k) \rangle$. The local OAM operators operate independently on the localized orbitals of each strand. For the full double-helix basis, the expectation values require evaluating the inter-orbital coherences across both structural backbones. The three spatial orbital textures are explicitly computed as:
\begin{align}
\langle L_r \rangle_\nu &= -2\hbar \text{Im} \left[ u_{1,\phi}^* u_{1,z} + u_{2,\phi}^* u_{2,z} \right], \label{eq:Lr} \\
\langle L_\phi \rangle_\nu &= -2\hbar \text{Im} \left[ u_{1,z}^* u_{1,r} + u_{2,z}^* u_{2,r} \right], \label{eq:Lphi} \\
\langle L_z \rangle_\nu &= -2\hbar \text{Im} \left[ u_{1,r}^* u_{1,\phi} + u_{2,r}^* u_{2,\phi} \right].
\label{eq:Lz}
\end{align}
The textures were obtained numerically and are depicted in Fig.~\ref{fig:textures}.
Figure~\ref{fig:textures} shows that inter-strand coupling generates additional orbital structure beyond the isolated single helix. In particular, panel (a) displays a finite radial texture $\langle L_r\rangle$. Although smaller than the azimuthal and longitudinal components, it is a sensitive signature of the incompatible phases introduced by asymmetric inter-strand hybridization, as discussed below.

The band-specific behavior in panels (b) and (c) reflects the hybridization strength of the specific transverse modes, showing enhanced structure near avoided crossings. The six band-resolved textures reflect the lifted degeneracies and hybridization of the asymmetric double helix.

\subsection{Symmetry Origin of the Radial Atomic OAM}

To clarify the origin of the nonzero radial component $\langle L_r\rangle$ in the
double-helix geometry, it is useful to consider first a symmetric reference configuration
defined by $\varphi_0=\pi$ and $z_0=0$. In this limit, the inter-strand cross-couplings
$t_{\perp,r\phi}$, $t_{\perp,rz}$, and $t_{\perp,\phi z}$ vanish. For the basis ordering
$(p_{1,r},p_{1,\phi},p_{1,z},p_{2,r},p_{2,\phi},p_{2,z})$, the diagonal unitary
transformation
$U_6=\operatorname{diag}(i,1,1,i,1,1)$
then transforms the full $6\times6$ Bloch Hamiltonian into a real matrix. Its eigenvectors
may consequently be chosen real in this representation. Since the radial orbital angular
momentum is determined by the relative phase between the $p_\phi$ and $p_z$ amplitudes,
\[
\langle L_r\rangle_\nu
=
-2\hbar\,
\operatorname{Im}
\left(
u_{1,\phi}^{*}u_{1,z}
+
u_{2,\phi}^{*}u_{2,z}
\right),
\]
the radial orbital texture vanishes in this symmetric reference limit.

Departing from this symmetric reference configuration introduces incompatible phase constraints through the inter-strand hybridizations. The major--minor groove asymmetry $\varphi_0\neq\pi$ generates the transverse cross-coupling $t_{\perp,r\phi}$, while a finite vertical shear $z_0\neq0$ activates the additional out-of-plane couplings $t_{\perp,rz}$ and $t_{\perp,\phi z}$. These terms are, in general, incompatible with the orbital rephasing that renders the intra-strand chiral hybridizations real. Consequently, the full Hamiltonian cannot, in general, be made simultaneously real by a diagonal rephasing of the orbital basis. The eigenvectors may therefore acquire nontrivial relative phases between their orbital components. In particular, the coherence $u_{\mu,\phi}^{*}u_{\mu,z}$ can acquire a finite imaginary part, producing the nonzero radial orbital texture observed in Fig.~\ref{fig:textures}.
Both the groove asymmetry and the vertical shear can therefore
contribute to the phase structure responsible for the radial atomic
texture; disentangling their individual quantitative contributions is
beyond the scope of the present work.

\section{Atomic OAM Transport and Response}
\label{sec:atomic_response}

Throughout this section, $L_\alpha$ denotes the local atomic OAM defined in Sec.~\ref{sec:atomic_oam}. Equilibrium currents and field-induced responses of this contribution are treated separately from the itinerant OAM introduced later in Sec.~\ref{sec:itinerant_oam}.

\subsection{Equilibrium Atomic Orbital Current}

A conventional OAM current is not, in general, forbidden in equilibrium by time-reversal symmetry because both the band velocity and a time-reversal-odd OAM texture change sign under $k\rightarrow-k$. To rigorously examine these properties within the asymmetric B-DNA double helix, we project the microscopic current-density operator onto the individual fully hybridized Bloch bands. The orbital angular momentum current operator associated with a spatial component $\alpha = (r, \phi, z)$ is defined through the canonical anti-commutator relation \cite{Shi2006}:
\begin{equation}
\hat{J}^{(L_{\alpha})}(k) = \frac{1}{2}\left\{ \hat{v}(k), \hat{L}_{\alpha} \right\},
\end{equation}
where $\hat{v}(k) = \frac{1}{\hbar}\partial_{k}\mathcal{H}(k)$ is the one-dimensional velocity operator. Within the semiclassical adiabatic wave-packet approximation, in which a Bloch state moves with the band group velocity $v_{\nu}(k)=\hbar^{-1}\partial_kE_{\nu}(k)$ while carrying its local on-site orbital moment, we approximate the band-resolved conventional current as:
\begin{equation}
J_{\nu}^{(L_{\alpha})}(k) \simeq v_{\nu}(k)\langle\hat{L}_{\alpha}(k)\rangle_{\nu}.
\end{equation}
The reduction of the projected anticommutator current to this band-velocity--times--orbital-texture form, together with the assumptions entering the adiabatic approximation, was derived explicitly for the corresponding multiorbital single-helix problem in Ref.~\cite{CordovaBercheMedina2026}. We therefore use the same approximation here and focus on the new consequences of inter-strand hybridization and double-helix geometry.
By aggregating the individual band current densities weighted by the equilibrium Fermi-Dirac distribution function $f_{0}(E_{\nu}(k))$, the total macroscopic equilibrium orbital angular momentum current is expressed as:
\begin{equation}
J_{\text{eq}}^{(L_{\alpha})} = \sum_{\nu}\int_{\mathrm{BZ}}\frac{dk}{2\pi}f_{0}\left(E_{\nu}(k)\right)J_{\nu}^{(L_{\alpha})}(k).
\end{equation}
Under strict time-reversal symmetry, the energy dispersion relation must obey $E_{\nu}(-k) = E_{\nu}(k)$, which naturally forces the band group velocity to act as a strictly odd function of momentum, $v_{\nu}(-k) = -v_{\nu}(k)$. As demonstrated via the time-reversal symmetry properties of the real-space hopping parameters, the orbital angular momentum textures are similarly constrained to be perfectly odd functions under momentum inversion, $\langle\hat{L}_{\alpha}\rangle_{\nu}(-k) = -\langle\hat{L}_{\alpha}\rangle_{\nu}(k)$. Crucially, the product of these two odd quantities is an even function of momentum throughout the Brillouin zone. Because the equilibrium distribution $f_{0}(E(k))$ is also even, symmetry does not require the integral over the Brillouin zone to cancel. Consequently, time-reversal symmetry does not force the conventional equilibrium orbital-current integral to vanish: a finite persistent-like equilibrium OAM current is symmetry-allowed in the double helix. Because orbital angular momentum is not generally conserved in the presence of orbital hybridization, the conventional equilibrium current does not, by itself, imply an accumulation of OAM at the molecular boundaries; a complete continuity equation must also include the corresponding orbital-torque term.

\subsection{Linear Atomic Orbital Edelstein Response}
\label{LinearAtomicResp}

For the atomic OAM textures introduced in Sec.~\ref{sec:atomic_oam}, the unbiased double-helix Hamiltonian preserves time-reversal symmetry ($\mathcal{T}$). This symmetry mandates that the energy dispersion is strictly even, $E_\nu(k) = E_\nu(-k)$, while the intrinsic orbital angular momentum (OAM) textures are antisymmetric, $\langle L_\alpha \rangle_\nu(k) = -\langle L_\alpha \rangle_\nu(-k)$. Consequently, integrating these textures over the symmetric Brillouin zone confirms that the net OAM identically vanishes in thermal equilibrium.

To drive a nonzero macroscopic response, the field produces a nonequilibrium odd-in-k occupation correction; the static Hamiltonian need not itself lose TR symmetry. We apply a uniform longitudinal electric field $E_z$ and evaluate the nonequilibrium electron dynamics using the semiclassical Boltzmann Transport Equation (BTE) within the relaxation time approximation. For each band, the steady-state nonequilibrium distribution $f_\nu(k)$ obeys
\begin{equation}
-\frac{eE_z}{\hbar}\frac{\partial f_\nu}{\partial k}
=-\frac{f_\nu-f_{0,\nu}}{\tau},
\end{equation}
where $e>0$ is the elementary charge, $\tau$ is the phenomenological momentum-relaxation time, and $f_0(E)=[\exp((E-\mu)/(k_BT))+1]^{-1}$ is the equilibrium Fermi--Dirac distribution at chemical potential $\mu$ and temperature $T$, with $k_B$ the Boltzmann constant. We use $\mathrm{BZ}$ for the first Brillouin zone and expand $f_\nu=f_{0,\nu}+f_{1,\nu}+f_{2,\nu}+\mathcal O(E_z^3)$. To first order,
\begin{equation}
f_{1,\nu}(k)
=\frac{eE_z\tau}{\hbar}\frac{\partial f_{0,\nu}}{\partial k}
=eE_z\tau v_\nu(k)
\left.\frac{\partial f_0}{\partial E}\right|_{E=E_\nu(k)},
\end{equation}
where $v_\nu(k)=\hbar^{-1}\partial_kE_\nu(k)$ is the band group velocity. Because the equilibrium energy $E_\nu(k)$ is an even function of momentum, the band velocity $v_\nu(k)$ is strictly odd. Therefore, the derivative $\partial_k f_{0,\nu}(k)$ is odd, and hence $f_{1,\nu}(k)=-f_{1,\nu}(-k)$. We can now evaluate the macroscopic nonequilibrium responses at the first order. The linear orbital Edelstein response is defined as the expectation value of the OAM operator over the perturbed distribution:
\begin{equation}
    \delta \langle L_\alpha \rangle^{(1)} = \sum_{\nu=1}^6 \int_{-\pi/h_z}^{\pi/h_z} \frac{dk}{2\pi} \, \langle L_\alpha \rangle_\nu(k) f_{1,\nu}(k).
\end{equation}
In this expression, the integrand is the product of the odd OAM texture $\langle L_\alpha \rangle_\nu(k)$ and the odd distribution $f_{1,\nu}(k)$. The resulting integrand is therefore an even function of $k$, surviving the integration over the symmetric Brillouin zone and yielding a robust, nonzero orbital Edelstein polarization in the linear regime. Defining $\delta\langle L_{\alpha}\rangle^{(1)}=\chi_{L_{\alpha}}E_z$, the linear orbital Edelstein susceptibility is
\begin{equation}
\chi_{L_{\alpha}}=
-e\tau \sum_{\nu}\int_{\mathrm{BZ}}\frac{dk}{2\pi}\,
v_{\nu}(k)\,\langle \hat L_{\alpha}\rangle_{\nu}(k)
\left(-\frac{\partial f_{0}}{\partial E}\right)_{E=E_{\nu}(k)}.
\end{equation}

\begin{figure}[t]
    \centering
    \includegraphics[width=0.9\linewidth]{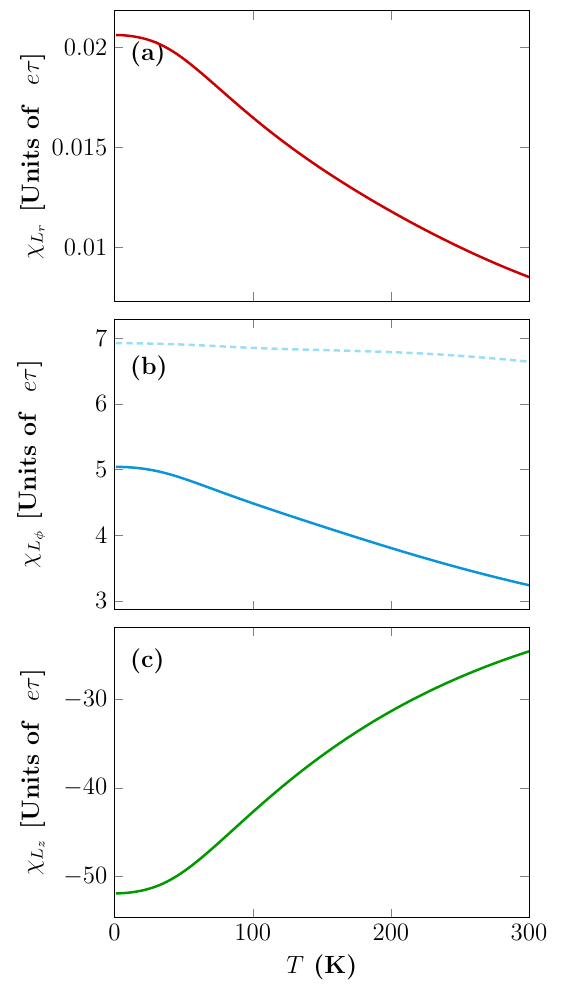}
    \caption{Temperature dependence of the linear atomic Edelstein susceptibility $\chi_{L_\alpha}$ at the chemical potentials that maximize the response. The different components reflect distinct orbital sectors and their thermal sampling near avoided crossings. (a) Radial response. (b) Azimuthal response near $\mu=1$ eV (solid line) and near $\mu=-1$ eV (dashed line). The azimuthal component is appreciable in two distinct energy windows. (c) Longitudinal response.}
    \label{fig:edelstein_response}
\end{figure}

Figure~\ref{fig:edelstein_response} shows the temperature dependence
of the three components of the linear atomic orbital Edelstein
susceptibility. For all components, the absolute magnitude of the
response decreases monotonically with increasing temperature, as
expected from the progressive thermal broadening of the Fermi surface.
The different decay scales reflect the distinct orbital sectors sampled
at the corresponding chemical potentials.
The latter are held fixed throughout each temperature sweep.
The values used for the radial, azimuthal, and longitudinal responses
are, respectively, $\mu_r = 1.02~\mathrm{eV},
\mu_{\phi,-}=-0.99~\mathrm{eV},
\mu_{\phi,+}=1.02~\mathrm{eV},
\mu_z=1.02~\mathrm{eV}$.
Two azimuthal curves are shown because sizeable responses occur in two
separate energy regions, one near each of the two quoted chemical
potentials.

In contrast, we evaluate the longitudinal nonequilibrium orbital current density $J_{L_\alpha}$ generated in this same linear regime:
\begin{equation}
    J_{L_\alpha}^{(1)} = \sum_{\nu=1}^6 \int_{-\pi/h_z}^{\pi/h_z} \frac{dk}{2\pi} \, v_\nu(k) \langle L_\alpha \rangle_\nu(k) f_{1,\nu}(k).
\end{equation}
Here, the integrand consists of the product of three odd functions: the velocity $v_\nu(k)$, the OAM texture $\langle L_\alpha \rangle_\nu(k)$, and the distribution correction $f_{1,\nu}(k)$. The product of three odd functions remains strictly odd. When integrated over the symmetric limits of the Brillouin zone, this term exactly cancels out. Therefore, parity dictates that while the applied electric field successfully generates a finite Edelstein OAM polarization (a static accumulation of angular momentum), the continuous nonequilibrium orbital current completely vanishes at the first order ($J_{L_\alpha}^{(1)} = 0$).

\subsection{Second-Order Atomic OAM Response}

Because the first-order field-induced correction to the orbital current
vanishes by parity, the leading nonequilibrium correction appears at
second order in the electric field. To fully characterize the transport behavior of the helical channel, we extend the semiclassical BTE formalism to the second order in the electric field ($E_z^2$). Recursively inserting $f_1(k)$ into the transport equation allows us to capture the second-order nonequilibrium correction driven by the higher-order gradients of the distribution function:
\begin{equation}
f_{2,\nu}(k)
=\frac{eE_z\tau}{\hbar}\frac{\partial f_{1,\nu}}{\partial k}
=\left(\frac{eE_z\tau}{\hbar}\right)^2
\frac{\partial^2 f_{0,\nu}}{\partial k^2}.
\end{equation}

Parity constraints strictly dictate which second-order macroscopic observables are permissible. Because $\partial_k f_0(k)$ is an odd function, its momentum derivative $\partial_k^2 f_0(k)$ must be strictly even. Consequently, the second-order distribution $f_{2,\nu}(k)$ is even in momentum. Applying this to the second-order OAM polarization, we find:
\begin{equation}
\begin{aligned}
\delta\langle L_\alpha\rangle^{(2)}
&=\left(\frac{eE_z\tau}{\hbar}\right)^2
\sum_{\nu=1}^6\int_{-\pi/h_z}^{\pi/h_z}\frac{dk}{2\pi}\\
&\quad\times\langle L_\alpha\rangle_\nu(k)
\frac{\partial^2 f_{0,\nu}(k)}{\partial k^2}=0.
\end{aligned}
\end{equation}
The integrand is an overall odd function of $k$, dictating that the orbital Edelstein response is forbidden at second order by parity ($\chi^{(2)}_{L_\alpha} = 0$). In contrast, we evaluate the nonequilibrium orbital current $J_{L_\alpha}$ at the second order. The orbital current operator scales as the product of velocity and angular momentum, yielding the integral:
\begin{equation}
\begin{aligned}
\delta\langle J_{L_\alpha}^{(2)}\rangle
&=\left(\frac{eE_z\tau}{\hbar}\right)^2
\sum_{\nu=1}^6\int_{-\pi/h_z}^{\pi/h_z}\frac{dk}{2\pi}\\
&\quad\times v_\nu(k)\langle L_\alpha\rangle_\nu(k)
\frac{\partial^2 f_{0,\nu}(k)}{\partial k^2}.
\end{aligned}
\end{equation}
Because the product of velocity and OAM texture is strictly even, the total integrand is an even function of momentum and survives the Brillouin zone integration. To evaluate this rigorously and avoid severe numerical artifacts from directly differentiating the sharp Fermi-Dirac distribution, we integrate by parts twice over the periodic Brillouin zone. Since $v_\nu(k)$, $\langle L_\alpha \rangle_\nu(k)$, and $f_{0,\nu}(k)$ are periodic at the zone boundaries, the boundary terms vanish. Defining the band current kernel $j_{\alpha,\nu}(k)\equiv v_\nu(k)\langle L_\alpha\rangle_\nu(k)$, we obtain the numerically stable form
\begin{equation}
\begin{aligned}
J_{L_\alpha}^{(2)}
&=\left(\frac{eE_z\tau}{\hbar}\right)^2
\sum_{\nu=1}^6\int_{-\pi/h_z}^{\pi/h_z}\frac{dk}{2\pi}\\
&\quad\times f_{0,\nu}(k)
\frac{\partial^2 j_{\alpha,\nu}(k)}{\partial k^2}\neq0.
\end{aligned}
\end{equation}
This form shows that the second-order orbital current depends directly on the momentum-space curvature of the orbital-velocity product within the occupied states. The field-induced orbital current to second order is $\langle J_{L_\alpha}^{(2)}\rangle=\sigma_{L_\alpha}^{(2)}E_z^2$, so the conductivity is given by:

\begin{equation}
\sigma_{L_\alpha}^{(2)}
=\left(\frac{e\tau}{\hbar}\right)^2
\sum_\nu\int_{\mathrm{BZ}}\frac{dk}{2\pi}\,
f_{0,\nu}(k)\frac{\partial^2 j_{\alpha,\nu}(k)}{\partial k^2}.
\end{equation}

\begin{figure}[t]
    \centering
    \includegraphics[width=0.9\linewidth]{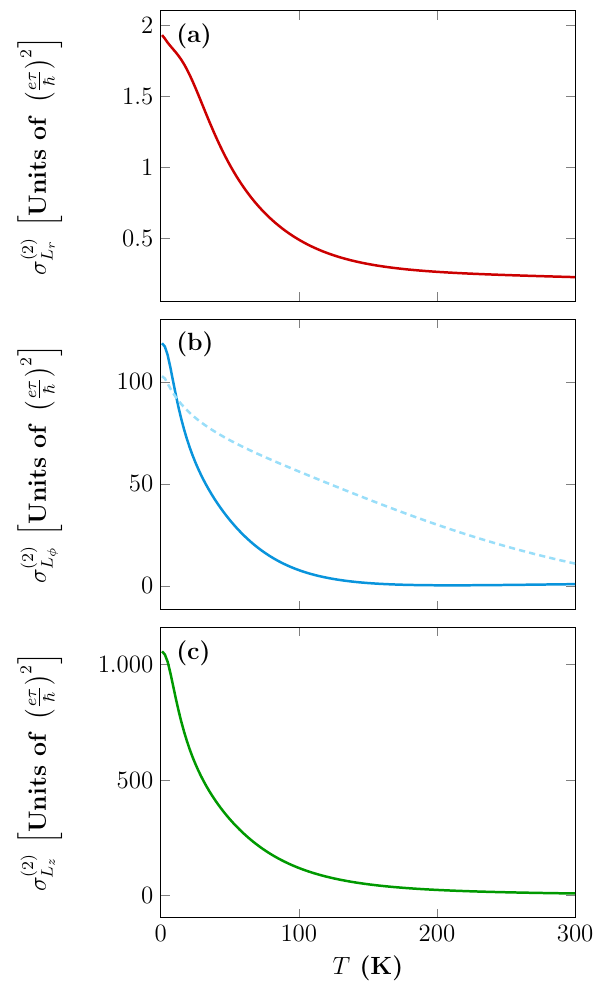}
    \caption{Temperature dependence of the second-order orbital current conductivity $\sigma^{(2)}_{L_\alpha}$ for (a) radial, (b) azimuthal (solid: $\mu = 1.14~\mathrm{eV}$, dashed: $\mu = -0.95~\mathrm{eV}$), and (c) longitudinal components. As in Fig.~\ref{fig:edelstein_response}, the azimuthal conductivity is appreciable in two chemical-potential windows.}
    \label{fig:current_response}
\end{figure}
Figure~\ref{fig:current_response} shows the corresponding quadratic conductivities. Similar to Fig.~\ref{fig:edelstein_response}, each conductivity has a chemical potential fixed so it maximizes the conductivity; for the radial, azimuthal, and longitudinal conductivities, the chemical potentials are, respectively (in eV), $\mu_r=1.13,\mu_{\phi_-}=-0.95,\mu_{\phi_+}=1.14,\mu_z=1.16$. All components decrease monotonically in magnitude as the temperature increases, reflecting the progressive thermal averaging of the momentum-resolved orbital-current structure.

\section{Itinerant Orbital Angular Momentum}
\label{sec:itinerant_oam}

Having completed the analysis of the local atomic orbital angular momentum, we now turn to the itinerant contribution. Both quantities originate from the same multiorbital tight-binding Hamiltonian, but they describe physically distinct forms of electronic circulation. The atomic contribution is associated with the intra-site orbital operator acting within the local $p$-orbital basis and is therefore determined by coherences between the amplitudes $(p_r,p_\phi,p_z)$ on a given site. The itinerant contribution instead originates from the motion of the electronic wave packet between distinct lattice sites and depends explicitly on the spatial geometry of the hopping network.

This distinction is particularly relevant for a double helix. In the local atomic sector, the second strand modifies the orbital hybridization and allows a radial OAM component that is absent in the corresponding isolated single-helix model. In the itinerant sector, the second strand changes the connectivity of the lattice itself: nearest-neighbor propagation along the two strands together with inter-strand hopping generates elementary closed circulation paths that are absent in a nearest-neighbor single helix. As shown below, these inter-strand loops possess a radial vector area in the co-rotating frame and therefore provide a new geometrical channel for a local radial itinerant OAM.

\subsection{Modern-theory definition of itinerant orbital angular momentum}

The position operator is not well defined for extended Bloch states in a periodic system, and the itinerant circulation cannot therefore be evaluated directly from the naive operator $\mathbf{r}\times\mathbf{p}$. The modern theory of orbital magnetization~\cite{xiao2005,Gobel2023ChiralOrbitronics} provides an origin-independent description in terms of the intrinsic orbital magnetic moment carried by a Bloch wave packet. For a cell-periodic Bloch eigenstate $|u_n(\mathbf{k})\rangle$ with energy $E_n(\mathbf{k})$, the intrinsic orbital magnetic moment is
\begin{equation}
\mathbf{m}^{\mathrm{it}}_n(\mathbf{k})
=
-\frac{e}{2\hbar}
\operatorname{Im}
\left[
\left\langle
\nabla_{\mathbf{k}}u_n
\right|
\times
\left(
H(\mathbf{k})-E_n(\mathbf{k})
\right)
\left|
\nabla_{\mathbf{k}}u_n
\right\rangle
\right],
\label{eq:modern_orbital_moment}
\end{equation}

where $e>0$ denotes the elementary charge. The corresponding itinerant orbital angular momentum is related to the orbital magnetic moment through
\begin{equation}
\mathbf{m}^{\mathrm{it}}_n
=
-\frac{e g_L}{2m_e}
\mathbf{L}^{\mathrm{it}}_n,
\label{eq:gyro_itinerant}
\end{equation}
where $m_e$ is the electron mass and $g_L=1$ for pure orbital motion. In this section, $H$ denotes the full six-band Bloch Hamiltonian $\mathcal H$ of Eq.~\eqref{eq:H6x6}. Using $k\cdot p$ perturbation theory,
\begin{equation}
\left|
\frac{\partial u_n}{\partial k_\beta}
\right\rangle
=
\sum_{m\neq n}
|u_m\rangle
\frac{
\langle u_m|
\partial_{k_\beta}H
|u_n\rangle
}{
E_n-E_m
},
\end{equation}
the three Cartesian components of the itinerant OAM may be written as
\begin{equation}
L^{\mathrm{it}}_{n,\alpha}(\mathbf{k})
=
\frac{2m_e}{g_L\hbar}
\sum_{m\neq n}
\frac{
\operatorname{Im}
\left[
\langle u_n|
\partial_{k_\beta}H
|u_m\rangle
\langle u_m|
\partial_{k_\gamma}H
|u_n\rangle
\right]
}{
E_m(\mathbf{k})-E_n(\mathbf{k})
},
\label{eq:itinerant_oam_vector}
\end{equation}
where $(\alpha,\beta,\gamma)$ denotes a cyclic permutation of $(x,y,z)$. Thus,
\begin{eqnarray}
    L^{\mathrm{it}}_{n,x}
&\;\longleftrightarrow\;&
\left(
\partial_{k_y}H,
\partial_{k_z}H
\right),\nonumber\\
L^{\mathrm{it}}_{n,y}
&\;\longleftrightarrow\;&
\left(
\partial_{k_z}H,
\partial_{k_x}H
\right),\nonumber\\
L^{\mathrm{it}}_{n,z}
&\;\longleftrightarrow\;&
\left(
\partial_{k_x}H,
\partial_{k_y}H
\right).
\end{eqnarray}
The longitudinal component $L_z^{\mathrm{it}}$ therefore represents only one component of the general itinerant orbital moment. The full modern-theory expression is vectorial, although the screw symmetry of an infinite helix implies that a spatially uniform transverse component cannot survive as a bulk laboratory-frame vector. This does not exclude a transverse itinerant texture that rotates together with the molecular frame.

\subsection{Bulk itinerant OAM and local circulating textures}

The distinction between a bulk transverse moment and a local transverse texture is essential in a helical system. Consider a local itinerant moment written in the co-rotating frame as
\begin{equation}
\mathbf{L}^{\mathrm{it}}_n
=
L_r^{\mathrm{it}}
\hat{\mathbf r}_n
+
L_\phi^{\mathrm{it}}
\hat{\boldsymbol\phi}_n
+
L_z^{\mathrm{it}}
\hat{\mathbf z}.
\label{eq:local_itinerant_texture}
\end{equation}
Under one screw operation, $n\rightarrow n+1$ and $\hat{\mathbf r}_{n+1}=R_z(\Delta\varphi)\hat{\mathbf r}_n$,
where $R_z(\Delta\varphi)$ denotes a rotation about the global $z$ axis; the same relation holds for $\hat{\boldsymbol\phi}_n$. Therefore $L_r^{\mathrm{it}}$ and $L_\phi^{\mathrm{it}}$ may remain finite and independent of $n$ in the co-rotating frame even though their Cartesian components rotate along the structure and vanish upon averaging over complete turns, $\overline{L_X^{\mathrm{it}}}
=
\overline{L_Y^{\mathrm{it}}}
=
0$.
The longitudinal component, by contrast, is referred to the fixed global helix axis and may survive the spatial average, $\overline{L_Z^{\mathrm{it}}}
\neq 0$.

The modern-theory expression in Eq.~\eqref{eq:itinerant_oam_vector} determines the bulk itinerant orbital moment. To understand the spatial origin of possible local transverse components, it is useful to complement this reciprocal-space formulation by examining the circulating part of the bond-current network generated by the same tight-binding Hamiltonian.
This real-space construction is not presented as an additional contribution to the modern-theory OAM. Rather, it provides a local geometrical interpretation of the inter-site circulation underlying the itinerant response.

\subsection{Local transverse circulation enabled by double-helix connectivity}

The second strand introduces elementary closed inter-site paths that do not exist in the nearest-neighbor single helix. For an oriented cycle ${\cal C}$ carrying a circulating component $I_{\cal C}$, the associated itinerant OAM is
\begin{equation}
\mathbf L_{\cal C}^{\mathrm{it}}
=-\frac{2m_e}{e g_L}I_{\cal C}\mathbf A_{\cal C},
\qquad
\mathbf A_{\cal C}=\frac{1}{2}\sum_{(ij)\in{\cal C}}
\mathbf R_i\times\mathbf R_j,
\label{eq:cycle_oam}
\end{equation}
where $\mathbf A_{\cal C}$ is the oriented vector area. Equation~\eqref{eq:cycle_oam} is a geometrical interpretation of the same inter-site circulation described by the modern-theory expression, not an additional contribution to it.

Appendix~\ref{app:itinerant_loops} shows that the elementary double-helix plaquette
$(1,n)\!\to\!(1,n+1)\!\to\!(2,n+1)\!\to\!(2,n)\!\to\!(1,n)$ has a vector area strictly radial, i.e., collinear with the local co-rotating radial direction. Hence a band state carrying circulation around this path has a local radial itinerant OAM. Time-reversed partners carry opposite circulation and cancel in equilibrium, whereas the odd-in-$k$ occupation shift induced by a longitudinal field makes a first-order radial response symmetry-allowed. We use this construction only to establish the transverse channel geometrically; its magnitude is not evaluated here. The explicit numerical calculation below is restricted to the longitudinal component, whose dominance in the present nearest-neighbor parameterization is discussed in Appendix~\ref{app:itinerant_dominance}.

\subsection{Itinerant OAM of the Double Helix}
\label{subsec:itinerant_double_helix}

We now apply the general modern-theory expression derived above to the
full six-band tight-binding Hamiltonian of the asymmetric double helix.
Although the itinerant orbital angular momentum is in general a vector
quantity, in the present work we evaluate explicitly its longitudinal
component $L_z^{\mathrm{it}}(k)$. As justified by the connectivity and
hopping hierarchy discussed in Appendix~\ref{app:itinerant_dominance},
this component is expected to provide the dominant itinerant contribution
for the nearest-neighbor double-helix model considered here.

The Bloch Hamiltonian of the molecular system is parameterized by the longitudinal crystal momentum $k\equiv k_z$. Nevertheless, the
longitudinal orbital moment is determined by the transverse position-weighted velocity matrix elements entering
Eq.~\eqref{eq:itinerant_oam_vector}. These matrix elements are fixed by
the full three-dimensional real-space embedding of the hopping network.

To generate them without introducing additional physical Bloch quantum
numbers, we temporarily augment the longitudinal phase factor by
auxiliary transverse phase variables $k_x$ and $k_y$. These variables
act only as generators of the Cartesian bond-displacement factors
$d_x$ and $d_y$; they do not represent independent crystal momenta of
the quasi-one-dimensional molecule. Thus, for a hopping contribution
between two basis states $a$ and $b$, connected by the spatial
displacement $\mathbf{d}_{ab}$, we define the auxiliary extension
\begin{equation}
H_{ab}(\mathbf{k})
=
\sum_{\mathbf{d}}
t_{ab}(\mathbf{d})
e^{i\mathbf{k}\cdot\mathbf{d}},
\label{eq:H_3D_embedding}
\end{equation}
where the sum includes all hopping processes connecting the
corresponding basis states. The auxiliary extension reduces to the
physical quasi-one-dimensional Hamiltonian on the line
$\mathbf{k}=(0,0,k)$.

The transverse derivatives entering the orbital-moment formula are then
generating derivatives evaluated on this physical line,
\begin{equation}
\left[D_x(k)\right]_{ab}
\equiv
\left.
\frac{\partial H_{ab}(\mathbf{k})}{\partial k_x}
\right|_{k_x=k_y=0}
=
i\sum_{\mathbf{d}}
d_x\,
t_{ab}(\mathbf{d})
e^{ikd_z},
\label{eq:Dx_exact}
\end{equation}
and
\begin{equation}
\left[D_y(k)\right]_{ab}
\equiv
\left.
\frac{\partial H_{ab}(\mathbf{k})}{\partial k_y}
\right|_{k_x=k_y=0}
=
i\sum_{\mathbf{d}}
d_y\,
t_{ab}(\mathbf{d})
e^{ikd_z}.
\label{eq:Dy_exact}
\end{equation}
Equations~(\ref{eq:Dx_exact}) and (\ref{eq:Dy_exact}) emphasize that
these transverse generating derivatives are not independent
phenomenological parameters. They are fixed by the same
Slater--Koster hopping amplitudes used to construct $H(k)$, weighted by
the actual Cartesian displacements of the corresponding bonds. In this
sense, $D_x$ and $D_y$ are compact momentum-space generators of the
position-weighted transverse hopping matrix elements of the embedded
quasi-one-dimensional lattice. The three-dimensional geometry of the
helix therefore enters the itinerant orbital moment through the
physical spatial factors $d_x$ and $d_y$, without assigning the
molecule fictitious transverse translational symmetry.

The exact eigenvalue problem of the full $6\times6$ Hamiltonian is
\begin{equation}
H(k)|u_\nu(k)\rangle
=
E_\nu(k)|u_\nu(k)\rangle.
\end{equation} We define the transverse interband matrix elements
\begin{equation}
\mathcal{D}^{x}_{\nu m}(k)
=
\langle u_\nu(k)|
D_x(k)
|u_m(k)\rangle ,
\label{eq:Dx_band}
\end{equation}
and
\begin{equation}
\mathcal{D}^{y}_{\nu m}(k)
=
\langle u_\nu(k)|
D_y(k)
|u_m(k)\rangle .
\label{eq:Dy_band}
\end{equation}
The longitudinal itinerant orbital angular momentum of band $\nu$ is
then obtained directly from the modern-theory expression as
\begin{equation}
L_{z,\nu}^{\mathrm{it}}(k)
=
\frac{2m_e}{g_L\hbar}
\sum_{m\neq\nu}
\frac{
\operatorname{Im}
\left[
\mathcal{D}^{x}_{\nu m}(k)
\mathcal{D}^{y}_{m\nu}(k)
\right]
}{
E_m(k)-E_\nu(k)
},
\label{eq:Lz_it_double_helix}
\end{equation}
with $g_L=1$ for pure orbital motion. Equation~
(\ref{eq:Lz_it_double_helix}) is the expression used for the numerical
evaluation of the itinerant OAM texture shown below.

This form is particularly convenient computationally. At each momentum $k$, the full Hamiltonian $H(k)$ is diagonalized to obtain $E_\nu(k)$ and $|u_\nu(k)\rangle$, the matrices $D_x(k)$ and $D_y(k)$
are constructed from Eqs.~(\ref{eq:Dx_exact}) and
(\ref{eq:Dy_exact}), and the interband sum in
Eq.~\eqref{eq:Lz_it_double_helix} is evaluated for each of the six bands. No assignment of an exact bonding or antibonding quantum number is required.

The geometrical content of Eq.~\eqref{eq:Lz_it_double_helix} becomes apparent
by considering two hopping contributions entering the product of
transverse derivatives. Their displacement factors occur in the
antisymmetric combination $d_x d_y'-d_y d_x'=(\mathbf{d}\times\mathbf{d}'
)_z$.

The longitudinal itinerant OAM therefore probes oriented areas projected onto the transverse $xy$ plane. In a helix, successive intra-strand hopping vectors rotate around the molecular axis, so propagation along the molecular backbone naturally carries a circulating component about the global $z$ direction. The
modern-theory expression in Eq.~\eqref{eq:Lz_it_double_helix} is the
reciprocal-space representation of this helical itinerant
circulation.

Because the Hamiltonian contains no magnetic field and preserves
time-reversal symmetry, $H(-k)=H^{*}(k)$,
the itinerant OAM texture is odd under momentum reversal,
\begin{equation}
L_{z,\nu}^{\mathrm{it}}(-k)
=
-
L_{z,\nu}^{\mathrm{it}}(k).
\label{eq:Lz_it_TR}
\end{equation}
This relation provides a direct numerical check of the calculation.
In particular, the itinerant texture must vanish at the
time-reversal-invariant momentum $k=0$, and the equilibrium
Brillouin-zone average of $L_z^{\mathrm{it}}$ vanishes in the absence of an occupation imbalance.

Accordingly, we use Eq.~\eqref{eq:Lz_it_double_helix} to calculate the
six band-resolved longitudinal itinerant textures
$L_{z,\nu}^{\mathrm{it}}(k)$. We leave the transverse itinerant components as symmetry-allowed secondary channels, as their detailed quantitative evaluation lies beyond the scope of the present work.

\begin{figure}[t]
    \centering
    \includegraphics[width=0.9\linewidth]{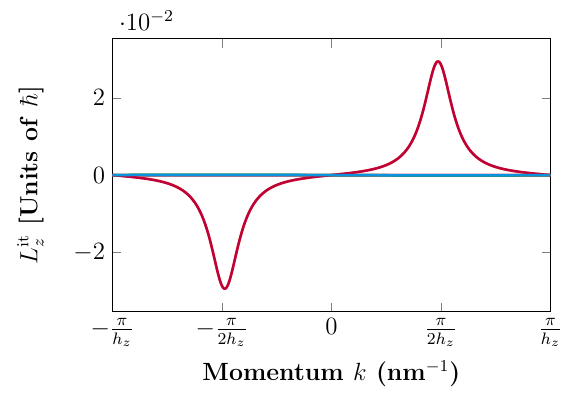}
    \caption{Longitudinal itinerant OAM texture $L_{z,\nu}^{\rm it}(k)$ obtained from Eq.~\eqref{eq:Lz_it_double_helix}. The colors label the same six bands as Fig.~\ref{fig:band_structure}; orbital and bonding assignments refer only to dominant eigenstate character. The texture is odd under $k\rightarrow-k$ as required by time-reversal symmetry.}
    \label{fig:iaomt_textures}
\end{figure}

Figure~\ref{fig:iaomt_textures} shows the six band-resolved longitudinal itinerant textures. A prominent feature is the localized resonant enhancement of $L_{z,\nu}^{\mathrm{it}}(k)$ occurring near $k \approx \pm \pi/(2h_z)$ exclusively for the hybridized transverse branches ($\nu=4, 5$), while the remaining bands are negligible. This sharp resonance peak arises directly from the interband energy denominator $[E_m(k) - E_\nu(k)]^{-1}$ in Eq.~\eqref{eq:Lz_it_double_helix}.  Despite the presence of these sharp interband resonances, the complete itinerant texture rigorously preserves global time-reversal symmetry ($\mathcal{T}$), adhering to the odd-parity relation from Eq.~\eqref{eq:Lz_it_TR}.

\subsection{Itinerant Orbital Linear Edelstein Response}

The nonequilibrium longitudinal itinerant accumulation is obtained from the same first-order occupation shift as the atomic response,
\begin{equation}
\delta\langle L_z^{\rm it}\rangle^{(1)}=
\sum_\nu\int_{\rm BZ}\frac{dk}{2\pi}
L_{z,\nu}^{\rm it}(k)f_{1,\nu}(k)
\equiv \chi_{L_z}^{\rm it}E_z.
\label{eq:it_Edelstein}
\end{equation}
Since both $L_{z,\nu}^{\rm it}(k)$ and $f_{1,\nu}(k)$ are odd in momentum, the integrand is even and the response is symmetry-allowed.

\begin{figure}[t]
    \centering
    \includegraphics[width=0.9\linewidth]{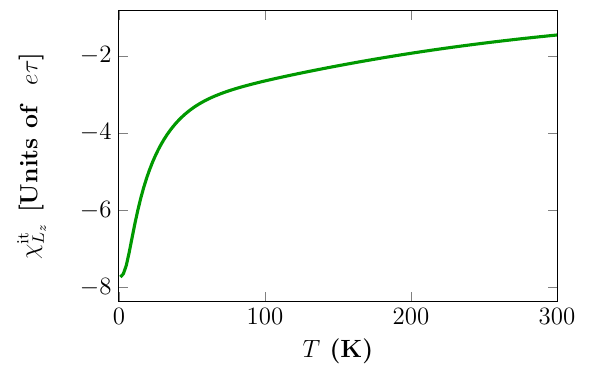}
    \caption{Temperature dependence of the longitudinal itinerant orbital Edelstein response susceptibility, $\chi_{L_z}^{\mathrm{it}}$ (in units of $e\tau$), evaluated at the optimal chemical potential $\mu \approx 1.14~\mathrm{eV}$.}
    \label{fig:ioamt_T_response}
\end{figure}

Figure~\ref{fig:ioamt_T_response} depicts the thermal evolution of the longitudinal itinerant orbital Edelstein susceptibility, $\chi_{L_z}^{\mathrm{it}}(T)$. The magnitude of the itinerant susceptibility decreases monotonically with temperature, approaching zero as thermal broadening progressively averages the sharply momentum-dependent itinerant texture.
In evaluating this response, the chemical potential is fixed at $\mu \approx 1.14\text{ eV}$ (near the energy where the bands $\nu=4,5$ exhibit an avoided crossing) rather than at the values of $\mu$ that maximize the local atomic Edelstein responses in Sec.~\ref{LinearAtomicResp}. While the local atomic orbital response $\chi_{L_z}^{\mathrm{at}}$ is driven by intra-site $p_r$--$p_\phi$ coherences concentrated primarily near the band centers, the itinerant response $\chi_{L_z}^{\mathrm{it}}$ is governed by the avoided crossing of the energy bands near $1.14\text{ eV}$.

\subsection{Second-order itinerant orbital-current response}
\label{sec:itinerant_second_order}

The longitudinal itinerant OAM texture can be placed on the same nonequilibrium footing as the atomic OAM discussed in Sec.~\ref{sec:atomic_response}. Within the same semiclassical intraband approximation, define the band-resolved itinerant current kernel
\begin{equation}
j^{\mathrm{it}}_{z,\nu}(k)\equiv v_\nu(k)L^{\mathrm{it}}_{z,\nu}(k).
\label{eq:j_it_kernel}
\end{equation}
Both factors are odd under $k\rightarrow-k$, so $j^{\mathrm{it}}_{z,\nu}(k)$ is even and its first-order field-induced correction vanishes. At second order, $f_{2,\nu}(k)$ is even.  The leading
field-induced itinerant OAM current is therefore
\begin{equation}
\delta J^{(2),\mathrm{it}}_{L_z}
=
\left(
\frac{eE_z\tau}{\hbar}
\right)^2
\sum_{\nu}
\int_{\mathrm{BZ}}
\frac{dk}{2\pi}\,
j^{\mathrm{it}}_{z,\nu}(k)
\frac{\partial^2 f_{0,\nu}(k)}{\partial k^2}.
\label{eq:J2_it_raw}
\end{equation}
The integrand is now even in momentum and the response is
symmetry-allowed.  Using the periodicity of the Brillouin zone and
integrating by parts twice gives
\begin{equation}
\delta J^{(2),\mathrm{it}}_{L_z}
=
\left(
\frac{eE_z\tau}{\hbar}
\right)^2
\sum_{\nu}
\int_{\mathrm{BZ}}
\frac{dk}{2\pi}\,
f_{0,\nu}(k)
\frac{\partial^2}{\partial k^2}
j^{\mathrm{it}}_{z,\nu}(k).
\label{eq:J2_it_ibp}
\end{equation}
Defining the quadratic itinerant orbital-current conductivity through
\begin{equation}
\delta J^{(2),\mathrm{it}}_{L_z}
=
\sigma^{(2),\mathrm{it}}_{L_z}E_z^2,
\end{equation}
we obtain
\begin{equation}
{
\sigma^{(2),\mathrm{it}}_{L_z}
=
\left(
\frac{e\tau}{\hbar}
\right)^2
\sum_{\nu}
\int_{\mathrm{BZ}}
\frac{dk}{2\pi}\,
f_{0,\nu}(k)
\frac{\partial^2}{\partial k^2}
\left[
v_{\nu}(k)L^{\mathrm{it}}_{z,\nu}(k)
\right],
}
\label{eq:sigma2_it}
\end{equation}
which is the itinerant counterpart of the second-order atomic orbital-current conductivity derived in Sec.~\ref{sec:atomic_response}.
The same parity argument also shows that the second-order itinerant OAM accumulation vanishes,
\begin{equation}
\delta \langle L^{\mathrm{it}}_z\rangle^{(2)}
=
\left(
\frac{eE_z\tau}{\hbar}
\right)^2
\sum_{\nu}
\int_{\mathrm{BZ}}
\frac{dk}{2\pi}\,
L^{\mathrm{it}}_{z,\nu}(k)
\frac{\partial^2f_{0,\nu}(k)}{\partial k^2}
=
0,
\end{equation}
because the itinerant texture is odd whereas
$\partial_k^2 f_0$ is even.  Thus the longitudinal itinerant sector
exhibits the same response hierarchy as the atomic one: a finite
linear Edelstein polarization but no linear field-induced orbital
current, followed at second order by a finite orbital-current response
but no quadratic OAM accumulation.

\section{Atomic versus Itinerant Orbital Angular Momentum}
\label{sec:atomic_vs_itinerant}
The longitudinal channel provides the common sector in which the two orbital contributions can be compared without identifying them. The atomic quantity $L_z^{\rm at}(k)$ measures on-site circulation within the local $p$-orbital basis, whereas $L_z^{\rm it}(k)$ measures inter-site wave-packet circulation through Eq.~\eqref{eq:Lz_it_double_helix}. This distinction also clarifies the relation to $s$-orbital helical models: such models retain the itinerant channel but, by construction, cannot describe the atomic OAM generated by coherence among local $p$ orbitals. In the present multiorbital double helix, both mechanisms coexist in the same Hamiltonian. Their momentum dependence and Edelstein susceptibilities therefore quantify physically distinct contributions to the total orbital response.

\begin{figure}[t]
    \centering
    \includegraphics[width=\linewidth]{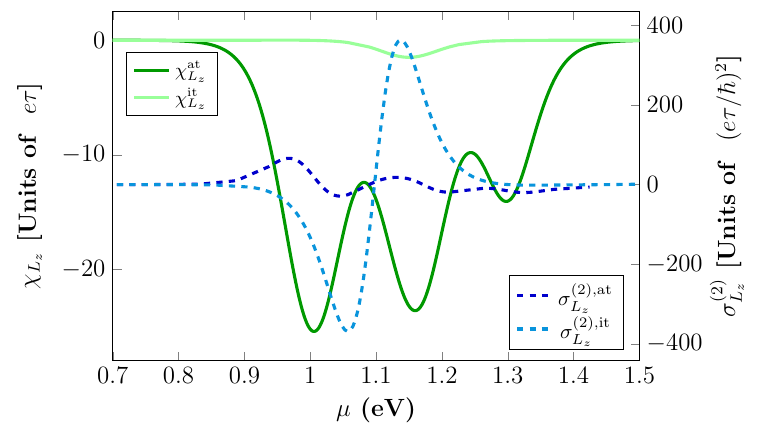}
    \caption{Longitudinal linear Edelstein susceptibilities ($\chi_{L_z}$, left axis, in units of $e\tau$) and second-order orbital-current conductivities ($\sigma^{(2)}_{L_z}$, right axis, in units of $(e\tau/\hbar)^2$) as functions of chemical potential at $T=300~\mathrm{K}$. Solid dark- and light-green curves denote the atomic and itinerant linear responses, respectively; dashed dark-blue and cyan curves denote the corresponding second-order orbital-current conductivities. The two orders exhibit distinct chemical-potential dependences: the linear response is predominantly atomic, whereas the itinerant quadratic conductivity develops pronounced resonant structure and sign changes, with its strongest enhancement in the $\mu\simeq1.0$--$1.15~\mathrm{eV}$ window.}
    \label{fig:combined_response_current}
\end{figure}

Figure~\ref{fig:combined_response_current} compares the longitudinal atomic and itinerant responses at $T=300$ K. The linear susceptibilities follow different chemical-potential dependences: the atomic contribution tracks local inter-orbital coherence, whereas the itinerant response is enhanced near interband avoided crossings. For the representative parameters, the atomic contribution dominates the linear response throughout most of the plotted range. The quadratic response displays a markedly different hierarchy: $\sigma^{(2),\mathrm{it}}_{L_z}$ develops pronounced resonant enhancements and sign changes in selected chemical-potential windows, particularly around $\mu\simeq 1.0$--$1.15~\mathrm{eV}$, where its magnitude substantially exceeds the atomic contribution. Thus, the relative importance of atomic and itinerant OAM depends not only on chemical potential but also on the response order.

The combined analysis also suggests a channel hierarchy. In the atomic sector, the calculated longitudinal texture is the largest, followed by the azimuthal component, whereas the radial component is much smaller but is a sensitive symmetry diagnostic of inter-strand phase frustration. In the itinerant sector, longitudinal winding is expected to dominate because it relies on the strong intra-strand network, whereas radial itinerant circulation requires weaker inter-strand transfer. We retain the transverse itinerant channels as symmetry-allowed possibilities but do not evaluate them quantitatively in this work.
   \label{fig:it_vs_at}

\section{Discussion and Conclusions}

We have developed a six-band tight-binding description of atomic and itinerant orbital angular momentum in an asymmetric double helix with characteristic B-form DNA geometry. Strand inequivalence and asymmetric inter-strand hopping prevent exact separation into bonding and antibonding sectors and, more importantly for the orbital response, generate relative phases among the local $p$-orbital amplitudes. These phases activate a radial atomic texture that is absent in the corresponding single-helix model, while the azimuthal and longitudinal atomic components remain finite.

The driven atomic response follows a simple parity hierarchy. Time-reversal symmetry makes the OAM texture odd in momentum, so a longitudinal electric field produces a finite first-order orbital Edelstein polarization but no first-order field-induced OAM current. At second order the parity is reversed: the OAM accumulation vanishes whereas a quadratic orbital-current response is allowed. The itinerant sector exhibits the same hierarchy in the longitudinal channel. Figure~\ref{fig:combined_response_current} compares the atomic and itinerant linear susceptibilities and quadratic current conductivities on common chemical-potential axes, showing that the relative importance of the two microscopic mechanisms depends on the response order and on the portion of the multiorbital spectrum being sampled.

The double helix also changes the geometry of itinerant circulation. Longitudinal $L_z^{\mathrm{it}}$ is generated by the ordinary helical winding along each strand and is evaluated explicitly with the modern Bloch-wave-packet expression. The second strand additionally creates elementary inter-strand plaquettes whose oriented vector area is purely radial in the co-rotating frame. This establishes a symmetry-allowed local radial itinerant channel distinct from the radial atomic texture: the former is an inter-site circulation effect, whereas the latter arises from on-site $p$-orbital coherence. Because the plaquette channel necessarily involves weaker inter-strand transfer, we expect it to be subleading to longitudinal itinerant winding except where interband resonances enhance it; its full quantitative evaluation is left for future work.

The present model deliberately isolates orbital physics. It uses nearest-neighbor hopping, a phenomenological relaxation time, and no explicit spin--orbit coupling, contacts, or dephasing mechanism. These ingredients will be required to connect the calculated OAM directly to spin-selective two-terminal observables. Within those limitations, the main result is robust: a double helix is not merely two copies of a single helical channel. Inter-strand coupling creates new atomic phase structure and new itinerant circulation topology, providing independent microscopic routes to geometry-controlled orbital polarization and nonlinear orbital transport in chiral molecular conductors.

\section*{Acknowledgments}

We acknowledge financial support from the Universidad San Francisco de Quito through the Poligrant POLI~41981. EV acknowledges support from the Doctoral program in Advanced Materials USFQ. We also acknowledge the use of AI-assisted tools for language refinement and discussion of theoretical directions during the preparation of this manuscript.

\appendix

\section{Microscopic Intra-Strand Parameters and Slater--Koster Expressions}
\label{app:slater_koster}

This appendix provides the explicit structural and algebraic elements governing the intra-strand tight-binding kinetics summarized in Sec.~\ref{section_tight-binding}. It also summarizes in Table~\ref{tab:bdna_parameters} all the B-DNA parameters used in the response and conductivity computations.
\begin{table}[t]
\centering
\caption{Geometric dimensions and representative tight-binding parameters used for the B-DNA-inspired double-helix model. The structural values follow characteristic B-form DNA scales, while the electronic parameters provide an effective multiorbital description consistent with model charge-transport energy scales \cite{hawke2010electronic, varela2016effective}.}
\label{tab:bdna_parameters}
\begin{ruledtabular}
\begin{tabular}{lc}
Parameter & Value \\
\hline
\multicolumn{2}{l}{\textbf{Geometric Dimensions}} \\
Radius ($a$) & $10.0$ \AA \\
Axial rise ($h_z$) & $3.4$ \AA \\
Azimuthal twist ($\Delta\varphi$) & $36^\circ$ \\
Groove asymmetry ($\varphi_0$) & $150^\circ$ \\
Vertical shear ($z_0$) & $0.15$ \AA \\
\hline
\multicolumn{2}{l}{\textbf{On-Site Energies: Strand 1}} (eV) \\
Radial ($\epsilon_{1,r}^0$) & $1.00$ \\
Azimuthal ($\epsilon_{1,\phi}^0$) & $1.15$ \\
Longitudinal ($\epsilon_{1,z}^0$) & $-1.00$ \\
\hline
\multicolumn{2}{l}{\textbf{On-Site Energies: Strand 2 ($\Delta\epsilon = 0.15$ eV)}} (eV) \\
Radial ($\epsilon_{2,r}^0 = \epsilon_{1,r}^0 + \Delta\epsilon$) & $1.15$ \\
Azimuthal ($\epsilon_{2,\phi}^0 = \epsilon_{1,\phi}^0 + \Delta\epsilon$) & $1.30$ \\
Longitudinal ($\epsilon_{2,z}^0 = \epsilon_{1,z}^0 + \Delta\epsilon$) & $-0.85$ \\
\hline
\textbf{Intra-Strand Hopping} & (meV) \\
Radial-radial ($t_{rr}$) & $-20.0$ \\
Azimuthal-azimuthal ($t_{\phi\phi}$) & $10.0$ \\
Longitudinal-longitudinal ($t_{zz}$) & $50.0$ \\
Chiral radial-azimuthal ($t_{r\phi}$) & $3.5$ \\
Chiral radial-longitudinal ($t_{rz}$) & $-14.3$ \\
Chiral azimuthal-longitudinal ($t_{\phi z}$) & $43.9$ \\
\hline
\textbf{Inter-Strand Hopping} & (meV) \\
Symmetric radial ($t_{\perp,rr}$) & $-18.0$ \\
Symmetric azimuthal ($t_{\perp,\phi\phi}$) & $-8.0$ \\
Symmetric longitudinal ($t_{\perp,zz}$) & $10.0$ \\
Parity-breaking radial-azimuthal ($t_{\perp,r\phi}$) & $-7.5$ \\
Parity-breaking radial-longitudinal ($t_{\perp,rz}$) & $-0.075$ \\
Parity-breaking azimuthal-longitudinal ($t_{\perp,\phi z}$) & $0.020$ \\
\end{tabular}
\end{ruledtabular}
\end{table}

Modeling the B-DNA single helical backbone with radius $a$, axial rise $h_z$, and an angular twist $\Delta\varphi$ between successive nucleotides, the intra-strand nearest-neighbor spatial distance $d$ is given by:
\begin{equation}
d = \sqrt{4a^2 \sin^2\left(\frac{\Delta\varphi}{2}\right) + h_z^2}.
\end{equation}
The unit bond vector $\hat{\mathbf{d}}_n = (\mathbf{R}_{\mu,n+1} - \mathbf{R}_{\mu,n})/d$ connecting adjacent sites along strand $\mu$ is projected onto the local cylindrical coordinate basis $(\hat{\mathbf{r}}, \hat{\bm{\phi}}, \hat{\mathbf{z}})$. These projections define the direction cosines $\alpha, \beta,$ and $\gamma$:
\begin{align}
\alpha &= -\hat{\mathbf{r}}_{\mu, n+1} \cdot \hat{\mathbf{d}}_{n}, \\
\beta &= \hat{\bm{\phi}}_{\mu, n+1} \cdot \hat{\mathbf{d}}_{n}, \\
\gamma &= \mathbf{\hat{z}} \cdot \hat{\mathbf{d}}_{n}.
\end{align}
By applying the Slater--Koster parameterization for $p$-orbital overlaps and defining the parameter $\Delta \equiv V_{pp\sigma} - V_{pp\pi}$, the microscopic intra-strand hopping elements are analytically evaluated as:
\begin{align}
t_{rr} &= V_{pp\pi} \cos\Delta\varphi - \Delta \alpha^2, \\
t_{r\phi} &= -V_{pp\pi} \sin\Delta\varphi + \Delta \alpha \beta, \\
t_{rz} &= \Delta \alpha \gamma, \\
t_{\phi\phi} &= V_{pp\pi} \cos\Delta\varphi + \Delta \beta^2, \\
t_{\phi z} &= \Delta \beta \gamma, \\
t_{zz} &= V_{pp\pi} + \Delta \gamma^2.
\end{align}
These matrix elements dictate the dispersion profiles and generate the momentum-dependent, imaginary odd-parity hybridization channels ($\gamma_{r\phi}(k)$, $\gamma_{rz}(k)$) and real even-parity coupling ($\gamma_{\phi z}(k)$) used to formulate the Bloch matrix blocks $H_\mu(k)$ in Eq.~\eqref{eq:H_k}. The complete form of the couplings $\gamma_{\alpha\beta}$ and the on-site terms $\epsilon_{\mu,\alpha}$ is given by
\begin{align}
\epsilon_{\mu,r}(k)&=\epsilon_{\mu,r}^0+2t_{rr}\cos(kh_z),\\
\epsilon_{\mu,\phi}(k)&=\epsilon_{\mu,\phi}^0+2t_{\phi\phi}\cos(kh_z),\\
\epsilon_{\mu,z}(k)&=\epsilon_{\mu,z}^0+2t_{zz}\cos(kh_z),\\
\gamma_{r\phi}(k)&=2t_{r\phi}\sin(kh_z),\\
\gamma_{rz}(k)&=2t_{rz}\sin(kh_z),\\
\gamma_{\phi z}(k)&=2t_{\phi z}\cos(kh_z).
\end{align}

\section{Geometrical Origin of Transverse Itinerant OAM}
\label{app:itinerant_loops}

This appendix gives the geometrical and symmetry analysis of the radial itinerant channel introduced in the main text.

\subsection{Single-helix winding and the new circulation topology of the double helix}

The distinction between the single and double helix is particularly transparent at the nearest-neighbor level. An isolated single helix possesses a directed helical propagation path and therefore supports itinerant circulation around the global molecular axis. This winding motion gives the longitudinal component $L_z^{\mathrm{it}}$ already identified in the single-helix problem.

However, a nearest-neighbor single strand contains no elementary closed inter-site plaquette. Its connectivity is simply
$
\cdots
\rightarrow
n-1
\rightarrow
n
\rightarrow
n+1
\rightarrow
\cdots$,
and consequently, there is no local nearest-neighbor circulation cycle analogous to the one introduced below. The double helix changes this topology even without extending the hopping range. Nearest-neighbor hopping along each strand, combined with inter-strand hopping, generates the elementary closed path
\begin{equation}
(1,n)
\rightarrow
(1,n+1)
\rightarrow
(2,n+1)
\rightarrow
(2,n)
\rightarrow
(1,n).
\label{eq:double_helix_loop}
\end{equation}
This plaquette represents a circulation channel that is absent in the corresponding nearest-neighbor single-helix model.

\subsection{Vector area of the elementary double-helix plaquette}

Using the coordinates defined for the two strands,
\begin{equation}
\mathbf R_{1,n}
=
a\cos(n\Delta\varphi)\hat{\mathbf X}
+
a\sin(n\Delta\varphi)\hat{\mathbf Y}
+
nh_z\hat{\mathbf Z},
\end{equation}
and
\begin{eqnarray}
\mathbf R_{2,n}
&=&
a\cos(n\Delta\varphi+\varphi_0)\hat{\mathbf X}
+
a\sin(n\Delta\varphi+\varphi_0)\hat{\mathbf Y}\nonumber\\
&+&
(nh_z+z_0)\hat{\mathbf Z},
\end{eqnarray}
the vector area of the plaquette in Eq.~\eqref{eq:double_helix_loop} is
\begin{eqnarray}
\mathbf A_n^{\square}
&=&
\frac{1}{2}
(
\mathbf R_{1,n}\times\mathbf R_{1,n+1}
+
\mathbf R_{1,n+1}\times\mathbf R_{2,n+1}\nonumber\\
&+&
\mathbf R_{2,n+1}\times\mathbf R_{2,n}
+
\mathbf R_{2,n}\times\mathbf R_{1,n}
).
\label{eq:plaquette_area_definition}
\end{eqnarray}
Introducing the azimuthal direction associated with the center of the plaquette,
\begin{equation}
\Theta_n
=
n\Delta\varphi
+
\frac{\Delta\varphi+\varphi_0}{2},
\end{equation}

and the corresponding co-rotating radial unit vector $
\hat{\mathbf r}_{c,n}
=
\cos\Theta_n\,\hat{\mathbf X}
+
\sin\Theta_n\,\hat{\mathbf Y}$,
the vector area reduces to
$\mathbf A_n^{\square}
=
A_r^{\square}
\hat{\mathbf r}_{c,n}$,
where
\begin{eqnarray}
A_r^{\square}
&=&
2a
\left[
z_0
\sin\left(\frac{\Delta\varphi}{2}\right)
\cos\left(\frac{\varphi_0}{2}\right)\right.\nonumber\\
&-& \left.
h_z
\cos\left(\frac{\Delta\varphi}{2}\right)
\sin\left(\frac{\varphi_0}{2}\right)
\right].
\label{eq:plaquette_radial_area_scalar}
\end{eqnarray}
For this elementary inter-strand plaquette, $A_\phi^{\square}=0,
A_z^{\square}=0$. Thus, the new elementary circulation path introduced by the second strand possesses a purely radial vector area in its natural co-rotating frame. A circulating electronic current around this plaquette can therefore generate a local radial itinerant OAM,
$\mathbf L_{n}^{\mathrm{it},\square}
=
L_r^{\mathrm{it},\square}
\hat{\mathbf r}_{c,n}$.
Because $\hat{\mathbf r}_{c,n}$ rotates by $\Delta\varphi$ from one unit cell to the next, this local radial itinerant component forms a helical texture and need not produce a nonzero uniform transverse moment in the laboratory frame.

\begin{figure}[t]
    \centering
    \includegraphics[width=\linewidth]{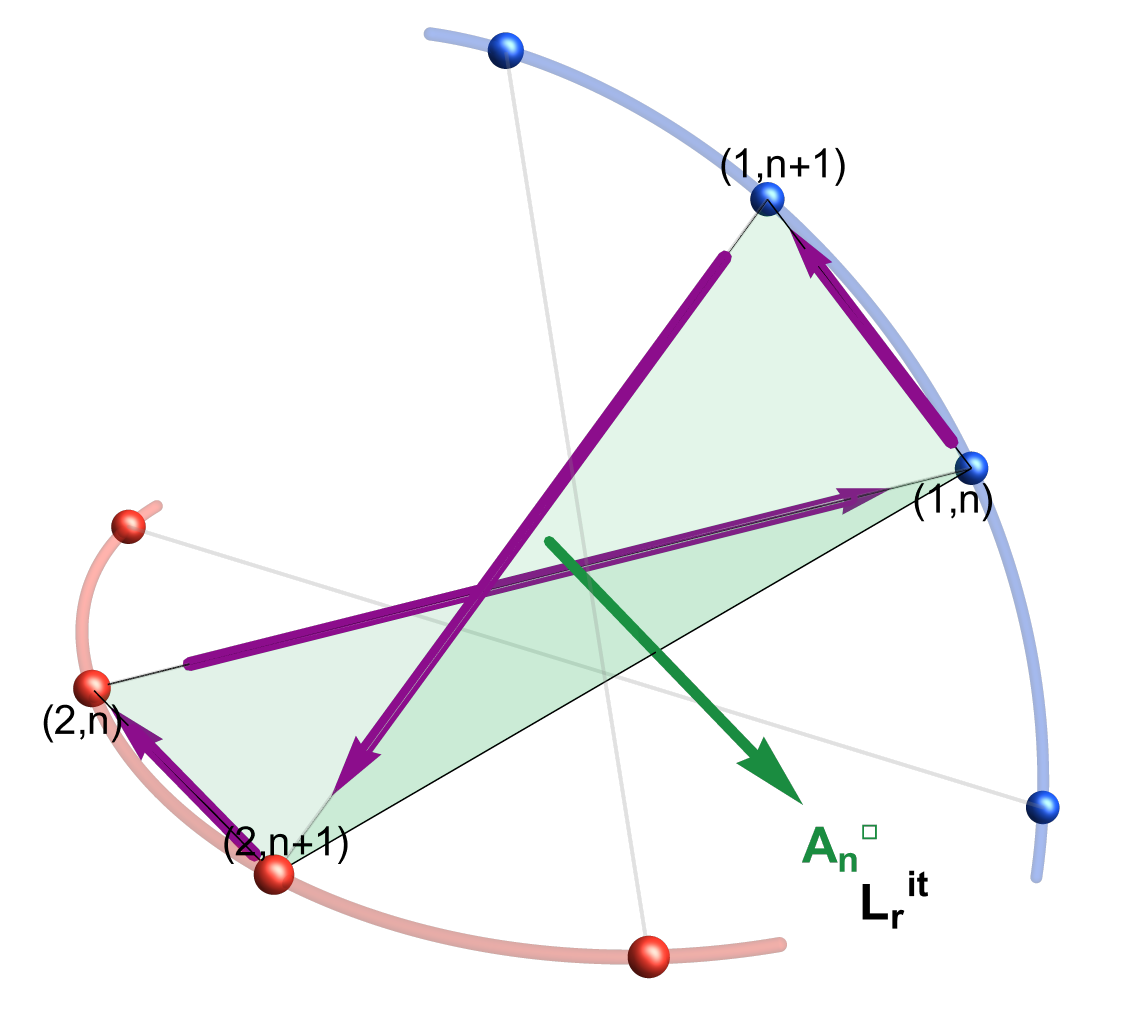}
\caption{
Elementary inter-strand plaquette of the double helix associated with
the radial itinerant orbital-angular-momentum channel. The directed
circulation follows the path
$(1,n)\rightarrow(1,n+1)\rightarrow(2,n+1)\rightarrow(2,n)
\rightarrow(1,n)$.
The corresponding oriented vector area
$\mathbf A_n^\square$ is collinear with the local co-rotating radial
direction $\hat{\mathbf r}_{c,n}$, demonstrating that this elementary
plaquette generates a purely radial itinerant OAM contribution $L_r^{it}$ (only its direction indicated).
The arrows indicate the oriented local circulation associated with this OAM channel; this circulating component is compatible with and superimposed on the direction of the longitudinal bias-driven transport.
}
\label{fig:itinerant_plaquette}
\end{figure}

Figure~\ref{fig:itinerant_plaquette} displays this elementary directed cycle and its radial oriented area.

The existence of the radial loop geometry does not require the major-minor groove asymmetry or the vertical shear. Indeed, in the symmetric geometrical limit $\varphi_0=\pi$ and $z_0=0$,
\begin{equation}
A_r^{\square}
=
-2ah_z
\cos\left(\frac{\Delta\varphi}{2}\right),
\end{equation}
which remains finite. The second strand, therefore, creates a radial itinerant circulation channel, even in the symmetric double helix. Structural and compositional asymmetries enter at the electronic level by modifying the band-resolved circulation amplitudes; time-reversal symmetry still enforces equilibrium cancellation, while a driven state may select a preferred circulation.

\subsection{Equilibrium cancellation and nonequilibrium selection of loop circulation}

A finite plaquette area does not by itself generate an orbital moment. In the absence of magnetic fields or other explicit time-reversal-symmetry breaking, the equilibrium system contains time-reversed electronic partners with opposite circulating currents. Denoting the band-resolved plaquette circulation by $I_{\square,\nu}(k)$, time-reversal symmetry requires
\begin{equation}
I_{\square,\nu}(-k)
=
-I_{\square,\nu}(k).
\label{eq:loop_current_TR}
\end{equation}
Because the equilibrium Fermi--Dirac distribution is even, $f_0[E_\nu(-k)]
=
f_0[E_\nu(k)],
$
the equilibrium average of the plaquette circulation vanishes,
\begin{equation}
I_{\square}^{\mathrm{eq}}
=
\sum_\nu
\int_{\mathrm{BZ}}
\frac{dk}{2\pi}
f_0[E_\nu(k)]
I_{\square,\nu}(k)
=
0.
\label{eq:loop_current_equilibrium}
\end{equation}
The radial plaquette geometry therefore defines an available itinerant OAM channel rather than an equilibrium radial orbital polarization.

A longitudinal nonequilibrium drive changes this conclusion. Within the relaxation-time Boltzmann description introduced previously, the distribution becomes
\begin{equation}
f_\nu(k)
=
f_{0,\nu}(k)
+
f_{1,\nu}(k)
+
\mathcal O(E_z^2),
\end{equation}
where $f_{1,\nu}(-k)=-f_{1,\nu}(k)$.
The applied field therefore preferentially populates one propagation direction over its time-reversed partner. In an asymmetric chiral double helix, the propagation direction can be correlated with a particular circulation sense around the inter-strand plaquette. If the band-resolved plaquette current obeys Eq.~(\ref{eq:loop_current_TR}), the nonequilibrium correction
\begin{equation}
\delta I_{\square}^{(1)}
=
\sum_\nu
\int_{\mathrm{BZ}}
\frac{dk}{2\pi}
f_{1,\nu}(k)
I_{\square,\nu}(k),
\label{eq:driven_loop_current}
\end{equation}
contains an even integrand and is therefore symmetry-allowed to remain finite.
Combining the driven plaquette circulation with Eq.~\eqref{eq:cycle_oam} and the radial vector area gives a local field-induced radial itinerant OAM.
This mechanism must be distinguished from spontaneous equilibrium circulation. The applied electric field does not act as an electromotive force around the microscopic plaquette. Instead, it generates an asymmetric occupation of propagating states. Chirality and structural or compositional asymmetry can correlate the direction of propagation with the local circulation carried by those states, so that the cancellation between opposite loop-current partners is incomplete away from equilibrium.

The physical sequence is therefore: double-helix loop geometry $\longrightarrow$
radial circulation channel, followed by nonequilibrium occupation imbalance
+ structural/compositional asymmetry $\longrightarrow$ preferred loop circulation $\longrightarrow \delta L_r^{\mathrm{it}}\neq0$.

Consequently, the second strand plays different roles in the atomic and itinerant sectors. For the atomic OAM, inter-strand hybridization modifies the relative phases of the local $p$-orbital amplitudes and activates the radial coherence entering $L_r^{\mathrm{at}}$. For the itinerant OAM, the second strand creates a new closed inter-site circulation path with a radial vector area. The two radial textures therefore arise from the same double-helix Hamiltonian but from physically distinct microscopic mechanisms.

\subsection{Connection with the longitudinal itinerant OAM}

The radial plaquette circulation described above is complementary to the longitudinal itinerant OAM associated with propagation around the global helix axis. We evaluate the latter explicitly in Sec.~\ref{subsec:itinerant_double_helix} from the Bloch-wave-packet orbital-moment expression, and it already exists for an isolated single helix.

At the nearest-neighbor level, the distinction between the two geometries can therefore be summarized as single-helix winding
$ \longrightarrow L_z^{\mathrm{it}}$, whereas double-helix inter-strand plaquette
$\longrightarrow L_r^{\mathrm{it}}$ under nonequilibrium circulation selection.

\subsection{Longitudinal versus transverse itinerant channels}
\label{app:itinerant_dominance}

The longitudinal itinerant component arises from propagation along the strong intra-strand hopping network: each nearest-neighbor step carries a transverse displacement as the backbone winds around the global $z$ axis. By contrast, the elementary radial channel necessarily uses the weaker inter-strand transfer amplitudes to close the plaquette. For the parameterization of Table~\ref{tab:bdna_parameters}, this connectivity provides a natural power-counting argument for retaining $L_z^{\mathrm{it}}$ as the dominant itinerant component in the explicit calculation. It is not an exact symmetry statement: interband denominators can enhance a transverse channel near avoided crossings, and longer-range hopping may generate additional circulation paths. The present model does not identify a comparably direct nearest-neighbor geometry for an azimuthal itinerant component.

\bibliography{bio2_corrected}

@article{reviewPaltiel,
  author  = {Naaman, Ron and Paltiel, Yossi and Waldeck, David H.},
  title   = {Chiral Molecules and the Spin Selectivity Effect},
  journal = {J. Phys. Chem. Lett.},
  volume  = {11},
  pages   = {3660--3666},
  year    = {2020},
  doi     = {10.1021/acs.jpclett.0c00474}
}

@article{bloom2024chiral,
  author  = {Bloom, Brian P. and Paltiel, Yossi and Naaman, Ron and Waldeck, David H.},
  title   = {Chiral Induced Spin Selectivity},
  journal = {Chem. Rev.},
  volume  = {124},
  pages   = {1950--1991},
  year    = {2024},
  doi     = {10.1021/acs.chemrev.3c00661}
}

@article{aiello2022chirality,
  author  = {Aiello, Clarice D. and Abendroth, John M. and Abbas, Muneer and Afanasev, Andrei and Agarwal, Shivang and Banerjee, Amartya S. and Beratan, David N. and Belling, Jason N. and Berche, Bertrand and Botana, Antia and Caram, Justin R. and Celardo, Giuseppe Luca and Cuniberti, Gianaurelio and Garcia-Etxarri, Aitzol and Dianat, Arezoo and Diez-Perez, Ismael and Guo, Yuqi and Gutierrez, Rafael and Herrmann, Carmen and Hihath, Joshua and Kale, Suneet and Kurian, Philip and Lai, Ying-Cheng and Liu, Tianhan and Lopez, Alexander and Medina, Ernesto and Mujica, Vladimiro and Naaman, Ron and Noormandipour, Mohammadreza and Palma, Julio L. and Paltiel, Yossi and Petuskey, William and Ribeiro-Silva, Jo{\~a}o Carlos and S{\'a}enz, Juan Jos{\'e} and Santos, Elton J. G. and Solyanik-Gorgone, Maria and Sorger, Volker J. and Stemer, Dominik M. and Ugalde, Jesus M. and Valdes-Curiel, Ana and Varela, Solmar and Waldeck, David H. and Wasielewski, Michael R. and Weiss, Paul S. and Zacharias, Helmut and Wang, Qing Hua},
  title   = {A Chirality-Based Quantum Leap},
  journal = {ACS Nano},
  volume  = {16},
  pages   = {4989--5035},
  year    = {2022},
  doi     = {10.1021/acsnano.1c01347}
}

@article{eckvahl2024detecting,
  author  = {Eckvahl, Hannah J. and Copley, Graeme and Young, Ryan M. and Krzyaniak, Matthew D. and Wasielewski, Michael R.},
  title   = {Detecting Chirality-Induced Spin Selectivity in Randomly Oriented Radical Pairs Photogenerated by Hole Transfer},
  journal = {J. Am. Chem. Soc.},
  volume  = {146},
  pages   = {24125--24132},
  year    = {2024},
  doi     = {10.1021/jacs.4c08706}
}

@article{varela2016effective,
  author  = {Varela, Solmar and Mujica, Vladimiro and Medina, Ernesto},
  title   = {Effective spin-orbit couplings in an analytical tight-binding model of {DNA}: Spin filtering and chiral spin transport},
  journal = {Phys. Rev. B},
  volume  = {93},
  pages   = {155436},
  year    = {2016},
  doi     = {10.1103/PhysRevB.93.155436}
}

@article{gutierrez2015spin,
  author  = {Gutierrez, R. and Cuniberti, G.},
  title   = {Spin-Dependent Effects in Helical Molecular Systems with Rashba-Like Spin-Orbit Interaction},
  journal = {Acta Phys. Pol. A},
  volume  = {127},
  pages   = {185--191},
  year    = {2015},
  doi     = {10.12693/APhysPolA.127.185}
}

@article{guo2014spin,
  author  = {Guo, Ai-Min and Sun, Qing-Feng},
  title   = {Spin-dependent electron transport in protein-like single-helical molecules},
  journal = {Proc. Natl. Acad. Sci. U.S.A.},
  volume  = {111},
  pages   = {11658--11662},
  year    = {2014},
  doi     = {10.1073/pnas.1407716111}
}

@article{yang2020detecting,
  author  = {Yang, Xu and van der Wal, Caspar H. and van Wees, Bart J.},
  title   = {Detecting Chirality in Two-Terminal Electronic Nanodevices},
  journal = {Nano Lett.},
  volume  = {20},
  pages   = {6148--6154},
  year    = {2020},
  doi     = {10.1021/acs.nanolett.0c02417}
}

@phdthesis{yang2020connecting,
  author  = {Yang, Xu},
  title   = {Connecting chirality and spin in electronic devices},
  school  = {University of Groningen},
  address = {Groningen, Netherlands},
  year    = {2020},
  doi     = {10.33612/diss.132019956}
}

@article{varela2024optical,
  author  = {Varela, Solmar and Gutierrez, Rafael and Cuniberti, Gianaurelio and Medina, Ernesto and Mujica, Vladimiro},
  title   = {Chiral spin selectivity and chiroptical activity in helical molecules},
  journal = {J. Chem. Phys.},
  volume  = {161},
  pages   = {114111},
  year    = {2024},
  doi     = {10.1063/5.0227365}
}

@article{Yoda2015HelicalMagnetization,
  author  = {Yoda, Taiki and Yokoyama, Takehito and Murakami, Shuichi},
  title   = {Current-induced Orbital and Spin Magnetizations in Crystals with Helical Structure},
  journal = {Sci. Rep.},
  volume  = {5},
  pages   = {12024},
  year    = {2015},
  doi     = {10.1038/srep12024}
}

@article{Yoda2018OrbitalEdelstein,
  author  = {Yoda, Taiki and Yokoyama, Takehito and Murakami, Shuichi},
  title   = {Orbital Edelstein Effect as a Condensed-Matter Analog of Solenoids},
  journal = {Nano Lett.},
  volume  = {18},
  pages   = {916--920},
  year    = {2018},
  doi     = {10.1021/acs.nanolett.7b04300}
}

@article{Go2018OrbitalTexture,
  author  = {Go, Dongwook and Jo, Daegeun and Kim, Changyoung and Lee, Hyun-Woo},
  title   = {Intrinsic Spin and Orbital Hall Effects from Orbital Texture},
  journal = {Phys. Rev. Lett.},
  volume  = {121},
  pages   = {086602},
  year    = {2018},
  doi     = {10.1103/PhysRevLett.121.086602}
}

@article{Cysne2025Orbitronics,
  author  = {Cysne, Tarik P. and Canonico, Luis M. and Costa, Marcio and Muniz, R. B. and Rappoport, Tatiana G.},
  title   = {Orbitronics in two-dimensional materials},
  journal = {npj Spintronics},
  volume  = {3},
  pages   = {39},
  year    = {2025},
  doi     = {10.1038/s44306-025-00103-1}
}

@article{varela2023spin,
  author  = {Varela, Solmar and Peralta, Mayra and Mujica, Vladimiro and Berche, Bertrand and Medina, Ernesto},
  title   = {Spin polarization induced by decoherence in a tunneling one-dimensional Rashba model},
  journal = {SciPost Phys. Core},
  volume  = {6},
  pages   = {044},
  year    = {2023},
  doi     = {10.21468/SciPostPhysCore.6.2.044}
}

@article{ozeri2023modification,
  author  = {Ozeri, Meital and Xu, Jiahui and Bauer, Gilad and Olde Olthof, Linde A. B. and Kimbell, Graham and Wittmann, Angela and Yochelis, Shira and Fransson, Jonas and Robinson, Jason W. A. and Paltiel, Yossi and Millo, Oded},
  title   = {Modification of Weak Localization in Metallic Thin Films Due to the Adsorption of Chiral Molecules},
  journal = {J. Phys. Chem. Lett.},
  volume  = {14},
  pages   = {4941--4948},
  year    = {2023},
  doi     = {10.1021/acs.jpclett.3c00702}
}

@article{mishra2020spin,
  author  = {Mishra, Suryakant and Mondal, Amit Kumar and Smolinsky, Eilam Z. B. and Naaman, Ron and Maeda, Katsuhiro and Nishimura, Tatsuya and Taniguchi, Tsuyoshi and Yoshida, Takumu and Takayama, Kokoro and Yashima, Eiji},
  title   = {Spin Filtering Along Chiral Polymers},
  journal = {Angew. Chem. Int. Ed.},
  volume  = {59},
  pages   = {14671--14676},
  year    = {2020},
  doi     = {10.1002/anie.202006570}
}

@article{GobelSchimpfMertig2025,
  author  = {G{\"o}bel, B{\"o}rge and Schimpf, Lennart and Mertig, Ingrid},
  title   = {Chirality-induced orbital Edelstein effect in an analytically solvable model},
  journal = {Phys. Rev. Res.},
  volume  = {7},
  pages   = {033180},
  year    = {2025},
  doi     = {10.1103/vpjm-ntbh}
}

@article{Liu2021ChiralityTopological,
  author  = {Liu, Yizhou and Xiao, Jiewen and Koo, Jahyun and Yan, Binghai},
  title   = {Chirality-driven topological electronic structure of {DNA}-like materials},
  journal = {Nat. Mater.},
  volume  = {20},
  pages   = {638--644},
  year    = {2021},
  doi     = {10.1038/s41563-021-00924-5}
}

@article{YatabeAkera2023,
  author  = {Yatabe, Yuya and Akera, Hiroshi},
  title   = {Coupling of orbital and spin polarizations to interatomic hopping in a helical atomic chain},
  journal = {J. Chem. Phys.},
  volume  = {159},
  pages   = {074701},
  year    = {2023},
  doi     = {10.1063/5.0156461}
}

@article{Georgiou2025Chirality,
  author  = {Georgiou, T. and Palma, J. L. and Mujica, V. and Schwartz, R. N. and Bouchard, L.-S.},
  title   = {Chirality-encoded molecular wavefunctions},
  journal = {J. Chem. Phys.},
  volume  = {163},
  pages   = {204104},
  year    = {2025},
  doi     = {10.1063/5.0293823}
}

@article{CordovaBercheMedina2026,
  author  = {Cordova, Danny and Berche, Bertrand and Medina, Ernesto},
  title   = {Orbital angular momentum textures and currents in a discrete helix: Equilibrium and linear response},
  journal = {J. Chem. Phys.},
  volume  = {165},
  pages   = {054309},
  year    = {2026},
  doi     = {10.1063/5.0343778}
}

@article{hawke2010electronic,
  author  = {Hawke, L. G. D. and Kalosakas, G. and Simserides, C.},
  title   = {Electronic parameters for charge transfer along {DNA}},
  journal = {Eur. Phys. J. E},
  volume  = {32},
  pages   = {291--305},
  year    = {2010},
  doi     = {10.1140/epje/i2010-10650-y}
}

@article{shui1998,
  author  = {Shui, Xiuqi and McFail-Isom, Lori and Hu, Gary G. and Williams, Loren Dean},
  title   = {The {B-DNA} Dodecamer at High Resolution Reveals a Spine of Water on Sodium},
  journal = {Biochemistry},
  volume  = {37},
  pages   = {8341--8355},
  year    = {1998},
  doi     = {10.1021/bi973073c}
}

@article{Shi2006,
  author  = {Shi, Junren and Zhang, Ping and Xiao, Di and Niu, Qian},
  title   = {Proper Definition of Spin Current in Spin-Orbit Coupled Systems},
  journal = {Phys. Rev. Lett.},
  volume  = {96},
  pages   = {076604},
  year    = {2006},
  doi     = {10.1103/PhysRevLett.96.076604}
}

@article{xiao2005,
  author  = {Xiao, Di and Shi, Junren and Niu, Qian},
  title   = {Berry Phase Correction to Electron Density of States in Solids},
  journal = {Phys. Rev. Lett.},
  volume  = {95},
  pages   = {137204},
  year    = {2005},
  doi     = {10.1103/PhysRevLett.95.137204}
}

@article{Gobel2023ChiralOrbitronics,
  author  = {Busch, Oliver and Mertig, Ingrid and G{\"o}bel, B{\"o}rge},
  title   = {Orbital Hall effect and orbital edge states caused by {$s$} electrons},
  journal = {Phys. Rev. Res.},
  volume  = {5},
  pages   = {043052},
  year    = {2023},
  doi     = {10.1103/PhysRevResearch.5.043052}
}
\end{document}